\documentclass[11pt,draftcls,onecolumn]{IEEEtran}
\usepackage[dvips]{epsfig}
\usepackage[dvips]{graphicx}
\usepackage{boxedminipage}
\usepackage{color}
\usepackage{amsfonts}
\usepackage{amssymb}
\usepackage{amsmath}
\usepackage{graphics,amsmath,amsfonts,amssymb,epsfig,latexsym,amsfonts,graphicx,amssymb}
\usepackage{eqparbox}

\makeatletter
\def\ps@headings{%
\def\@oddhead{\mbox{}\scriptsize\rightmark \hfil \thepage}%
\def\@evenhead{\scriptsize\thepage \hfil \leftmark\mbox{}}%
\def\@oddfoot{}%
\def\@evenfoot{}}
\makeatother 

\newcommand{\bfbeta}{{\mbox{\boldmath $\beta$}}}

\newcommand{\bfPhi}{{\mbox{\boldmath $\Phi$}}}

\def\BibTeX{{\rm B\kern-.05em{\sc i\kern-.025em b}\kern-.08em
    T\kern-.1667em\lower.7ex\hbox{E}\kern-.125emX}}

\begin{document}
\title{Distributed Uplink Resource Allocation in
Cognitive Radio Networks -- Part II: Equilibria and Algorithms for
Joint Access Point Selection and Power Allocation}
\author{Mingyi Hong, Alfredo Garcia and Jorge Barrera
\thanks{M. Hong, A. Garcia and J. Barrera are with the Department of Systems and Information Engineering,
University of Virginia, Charlottesville, VA} \thanks{Part of this
manuscript has been accepted by the Proceedings of IEEE INFOCOM
2011\cite{hong11_infocom}.}}

\maketitle
\begin{abstract}
The main objective of this two part paper is to formulate and
address the problem of distributed uplink resource allocation in
multi-carrier cognitive radio networks (CRN) with multiple Access
Points (APs). When the APs operate on non-overlapping spectrum
bands, such problem is essentially a joint spectrum decision and
spectrum sharing problem. In this network, the cognitive users (CUs)
are endowed with greater flexibility than the single AP network we
considered in the first part of the paper \cite{hong10m_japsa_1}:
they can optimize their uplink transmission rates by means of: 1)
association to a suitable AP and 2) sharing the set of channels that
belong to this AP with other CUs associated with this AP. Clearly
these two steps are interdependent, and our objective is to devise
suitable algorithms by which the CUs can perform these two steps in
a distributed and efficient fashion.

In the first part of this paper, we have studied solely the spectrum
sharing aspect of the above problem, and proposed algorithms for the
CUs in the single AP network to efficiently share the spectrum. In
this second part of the paper, we build upon our previous
understanding of the single AP network, and formulate the joint
spectrum decision and spectrum sharing problem in a multiple AP
network into a non-cooperative game, in which the feasible strategy
of a player contains a discrete variable (the AP/spectrum decision)
and a continuous vector (the power allocation among multiple
channels). The structure of the game is hence very different from
most non-cooperative spectrum management game proposed in the
literature. We provide characterization of the Nash Equilibrium (NE)
of this game, and present a set of novel algorithms that allow the
CUs to distributively and efficiently select the suitable AP and
share the channels with other CUs. Finally, we study the properties
of the proposed algorithms as well as their performance via
extensive simulations.

\end{abstract}
\section{Introduction}
\subsection{Motivation and Related Work}\label{subMotivation}

The objective of this two part paper is to provide the analytical
framework as well as the solutions to the joint AP selection and
power allocation problem in a CRN in the presence of multiple APs.
As mentioned in the first part of the paper, the need for such joint
optimization may arise in a CRN with multiple CUs and multiple APs,
for example, the IEEE 802.22 cognitive radio Wireless Regional Area
Network (WRAN) \cite{stevenson09}. In such network, a particular
geographical region may be served by multiple service providers
(SPs), or by multiple APs installed by a single SP \cite{acharya09}.
Consequently, the CUs, on top of being able to share the spectrum
offered by a particular SP/AP, also have the flexibility of deciding
on their SP/AP association. As suggested in \cite{apcan06} and
\cite{saraydar01}, it would be generally beneficial (in terms of
either system-wide or individual performance), compared with
traditional closest AP assignment strategy, to allow the users in
multiple AP networks to include the AP association as an additional
decision variable.


In this part of the paper, we consider a CRN with multiple CUs and
APs. The available spectrum is partitioned by the APs, and they
operate on non-overlapping spectrum bands. Each CU's objective is to
connect to a single AP for communication. The CUs can concurrently
use all the channels that belong to its associated AP for
transmission if desired, but different CUs interfere with each other
if they use the same channel. In the considered network, the CUs
first need to select an appropriate AP for communication, a task
that can be viewed as a spectrum decision task because the CUs are
essentially choosing a ``best" spectrum band in terms of
transmission rate. Then they face the spectrum sharing problem when
they try to dynamically allocate their communication power across
the channels that belong to the selected AP. Clearly for a {\it
fixed} system wide CU-AP association, the entire network is reduced
to multiple single AP sub-networks, and the (near-) optimal spectrum
sharing solution for each of the sub-network is studied in the first
part of this paper \cite{hong10m_japsa_1}. Although as we have
demonstrated that for each sub-network, our proposed spectrum
sharing scheme has the potential of maximizing the achievable sum
rate, in a multiple AP network, the system performance is inevitably
tied to the quality of the system wide CU-AP association decision as
well. A bad CU-AP association decision will result in unsatisfactory
system performance regardless of the underlying sharing scheme.
Consequently, the association problem and the sharing problem are
strongly interdependent, and in this part of the paper, we intend to
propose distributed and efficient algorithm for the CUs in the
network to carry out both of the tasks of AP association (spectrum
decision) and power allocation (spectrum sharing).

A related problem of joint cell selection/base station (BS)
association and power control has been addressed in
infrastructure-based cellular networks. \cite{hanly95} and
\cite{yates95b} are early works trying to tackle this problem in an
uplink spread spectrum cellular network. The objectives are to let
the users find a best site selection and power allocation tuple such
that all users' target signal to interference ratio (SIR) are met,
and each user's transmission power is minimized. The authors of
\cite{apcan06} and \cite{saraydar01} cast a similar problem (with an
objective to maximize individual power efficiency or minimizing
individual cost) into game theoretical frameworks, and propose
algorithms to find the Nash Equilibrium (NE) of the proposed games.
One of the most important differences between our work and the above
cited works is that the power allocation problems in these works are
essentially {\it scalar value} optimization problem: each user only
needs to decide on its {\it power level} once a BS is selected,
while in our work, individual power allocation is a {\it vector
optimization} problem as the CUs have the flexibility to use all the
channels that belong to a particular AP concurrently. This
fundamental difference makes the considered problem more complex,
hence the analytical frameworks provided by the above cited works
are not suitable for our problem. \cite{Meshkati06} is a recent work
using non-cooperative game theory to address the problem of
distributed energy-efficient power control in uplink multi-carrier
CDMA system. Similarly as in the above cited works, the solution
proposed by the authors mandates that the users choose a single
optimum channel as well as a {\it scalar} power level to transmit on
the selected channel. \cite{acharya09} is a recent work considering
the uplink dynamic spectrum sharing problem in a multi-carrier
multiple service provider CRN. The authors propose algorithms for
the users to select the size of the spectrum and the amount of power
for transmission. One important assumption of this work is that the
users can connect to multiple APs at the same time (we refer to such
network as multiple-connectivity network), an assumption that
simplifies the analysis significantly but may induce considerable
signaling overhead on the network side as well as hardware
implementation complexity on the cognitive device\footnote{In WLAN
literature, such network is also referred to as ``multi-homing"
network, see \cite{shakkottai07} and the reference therein.}. Even
such issues may be resolved in the future, our work, which analyzes
the single-connectivity network, can serve as a benchmark for
comparison between single-connectivity and multiple-connectivity
networks.

We also argue that the problem under consideration is in many
aspects more complicated than the traditional AP association
problems arise in the 802.11 WLAN network (for example,
\cite{kauffmann07}, \cite{bejerano07} and \cite{kumar05} and the
reference therein). Typically, AP association is aiming to optimize
different system performance metrics (throughput, fairness, etc),
and only simple individual throughput estimates within each AP are
used to update the current association profile. Indeed, in 802.11
WLAN network, the throughput of an individual AP with fixed number
of users and fixed physical bit rate can be approximated using
simple analytical formulae \cite{kumar07}, and this result has
greatly simplified the analysis of many work dealing with dynamic AP
association in WLAN, e.g., \cite{kumar05} and \cite{shakkottai07}.

We note here that the problem of how to dynamically perform the task
of both spectrum decision and spectrum sharing may arise in
different CRN configuration as well. Many of the current works
addressing the spectrum management problem in multi-channel
multi-user CRN focus only on the spectrum sharing aspect of the
problem. For example, in \cite{wang08}, \cite{pang09}, \cite{wu09},
a set of iterative water-filling \footnote{IWF is originally
proposed in \cite{yu02a} in the context of DSL network, and
subsequently applied to wireless network with vector multiple-access
channel \cite{yu04} and with gaussian interference channel in
\cite{luo06b}, \cite{scutari08a} \cite{scutari08b}.} (IWF) based
algorithms are proposed to find a distributed solution of power
allocation in multi-channel, multi-user CRN. One important
assumption underlying these works is that the CUs are able to use
{\it  all the channels simultaneously}. This assumption might not be
valid in the situation where the available spectrum is fragmented
due to licensed user activities and the CUs are equipped with
1-agile radio which can only use a single chunk of continuously
aligned channels at a time \footnote{See \cite{cao10} for discussion
of agile radios and the possibility of this scenario in actual CRN
implementations.}.  In this scenario, the CUs need to first decide
on {\it which chunk of channels} to use, and then make subsequent
power allocation decisions on the selected set of channels, i.e.,
the CUs are required to perform the task of joint spectrum decision
and spectrum sharing. It is our belief that our work can also serve
to shed some lights on providing solutions to the above problem, as
the network configuration considered in our work is sufficiently
similar to the configuration mentioned above.

\subsection{Contributions and Organization of This Work}
To the best of our knowledge, this is the first work that proposes
distributed algorithms to deal with joint AP selection and power
allocation problem in a multi-channel multi-AP CRN. We cast the
problem into a non-cooperative game framework, in which each CU's
objective is to maximize its own transmission rate, and its strategy
space is the union of a {\it discrete set} (the set of possible APs)
and a {\it multi-dimensional continuous set} (the set of feasible
power vectors). Although non-cooperative game theory has recently
been extensively applied to solve the resource allocation problem in
CRN (e.g., \cite{wang08}, \cite{wu09}, \cite{zhang08} and the
reference therein), our formulation is considerably different and
more involved because of such ``hybrid" nature of the strategy space
of the game. We analyze in detail the equilibrium solution of the
game, and develop a suite of algorithms with provable convergence
guarantees that enable the CUs to distributedly compute the
equilibrium solution.

We organize our paper as follows. In section \ref{secSystemModel},
we present the network model under consideration and formulate the
problem into a non-cooperative game. In section \ref{secNE}, we
analyze the properties of the equilibrium solution. In section
\ref{secAlgorithm}, we provide our main algorithm and its
convergence results. In section \ref{secExtension} and
\ref{secJJASPA}, we provide important extensions of the JASPA. We
present simulation results in section \ref{secSimluation} and
conclude the paper in section \ref{secConclusion}.

%

\section{Problem Formulation}\label{secSystemModel}
\subsection{Considered Network and Some Assumptions}
We consider the following cognitive network configuration. Suppose
there are a set $\{1,2,\cdots,N\}\triangleq\mathcal{N}$ of CUs, a
set $\{1,2,\cdots,K\}\triangleq\mathcal{K}$ of channels and a set
$\{1,2,\cdots,W\}\triangleq\mathcal{W}$ of APs in the network, and
we normalize the total available bandwidth to $1$. Each AP
$w\in\mathcal{W}$ is assigned with a subset of channels
$\mathcal{K}_w\subseteq\mathcal{K}$. We focus on the uplink scenario
where each CU wants to connect to one of the APs for transmission.
The
followings are our main assumptions of the network.\\
{\bf A-1)} Each CU $i$ is able to associate to all the APs, and each
AP covers entire area of the network.\\
{\bf A-2)} The APs covering the same area operate on non-overlapping
portions of the available spectrum.\\
{\bf A-3)} The set of spectrum can be used exclusively by
the CRN for a relative long period of time.\\
{\bf A-4)} Each CU can associate to a single AP at a time; it can
concurrently use all the channels of the associated AP, if
desired.\\
{\bf A-5)} Each AP is equipped with single-user receivers. Different
APs in the network do not compete with each other for revenue.

Assumption A-1) is made merely for ease of presentation, and our
work can be extended to the scenarios where different APs cover
different areas of the network, and where the CUs can only connect
to the subset of APs that cover them.

Assumption A-2) is commonly used when considering AP association
problems in WLAN (e.g., \cite{shakkottai07}), or the spectrum
sharing problem in cognitive network with multiple service providers
(e.g.,\cite{acharya09}). It is made to mitigate interference between
neighboring APs. It can be achieved either by 1) the APs agree
offline the partition of the spectrum \footnote{In the presence of
multiple SPs, such offline negotiation can be made possible by the
coordination of a spectrum clearing house, as suggested in
\cite{acharya09}.} or 2) the APs jointly run a distributed online
spectrum assignment algorithm similar to the ones proposed in
\cite{kauffmann07} to determine the best spectrum assignment. How to
determine the ``optimum" partition of the spectrum is out of the
scope of this paper. Assumption A-1) and A-2) imply that
$\mathcal{K}_w\bigcap\mathcal{K}_q=\emptyset,~\forall~w\ne q,
k,q\in\mathcal{W}$.



\subsection{System Model}
Let $\{|h_{i,w}(k)|^2\}_{k\in\mathcal{K}_w}$ be the set of power
gains from CU $i$ to AP $w$ on all its channels; Let
$\{n_w(k)\}_{k\in\mathcal{K}_w}$ be the set of environmental noise
powers on all channels for AP $w$; Let the $N\times 1$ vector
$\mathbf{a}$ denote the {\it association profile} in the network,
with its $i^{th}$ element $\mathbf{a}(i)=w$ indicating that CU $i$
is associated to AP $w$. Each CU $i$ is able to obtain {\it its own}
channel gains to all the APs,
$\{|h_{i,w}(k)|^2\}_{k\in\mathcal{K}_w, w\in\mathcal{W}}$, via
feedback from the APs, but it {\it does not} need to have the
knowledge of other CUs' channel gains in the network.

Let ${p}_{i,w}(k)$ represent the amount of power CU $i$ transmits on
channel $k$ when it is associated with AP $w$; Let
$\mathbf{p}_{i,w}=\left\{p_{i,w}(k)\right\}_{k\in\mathcal{K}_w}$ be
the { power profile} of CU $i$ when it is associated with AP $w$;
let $\mathbf{p}_{-i,w}$ be the joint power profiles of all the CUs
other than $i$ that is associated with AP $w$:
$\mathbf{p}_{-i,w}\triangleq\{\mathbf{p}_{j,w}\}_{j:j\ne i,
\mathbf{a}(j)=w}$. By construction, for all $w\in\mathcal{W}$, if
$w\ne\mathbf{a}(i)$, then $\mathbf{p}_{i,w}= \mathbf{0}$. The power
profiles of the CUs must satisfy the following two constraints (as
in \cite{hong10m_japsa_1}): 1) Total power constraints; 2)
Positivity constraints. As such, each CU's feasible power allocation
when it is associated with AP $w$ can be expressed
as:\vspace{-0.15cm}
\begin{align}
\vspace{-0.15cm}\mathcal{F}_{i,w}=\Big\{\mathbf{p}_{i,w}:\hspace{-0.1cm}\sum_{k\in\mathcal{K}_w}p_{i,w}(k)\le
\bar{p}_i,~p_{i,w}(k)\ge
0,~\forall~k\in\mathcal{K}_w\Big\}.\nonumber
\end{align}
Again assume that there is no interference cancelation performed at
the AP, then for a fixed AP association and power allocation
configuration, CU $i$'s uplink transmission rate (when it is
associated with AP $w$) can be expressed as follows:\vspace{-0.15cm}
\begin{align}
&\hspace{-0.25cm}R_i(\mathbf{p}_{i,w},\mathbf{p}_{-i,w};
w)\nonumber\\
&\hspace{-0.25cm}=\frac{1}{K}\hspace{-0.15cm}\sum_{k\in\mathcal{K}_w}\hspace{-0.12cm}
\log\hspace{-0.08cm}\Big(\hspace{-0.05cm}1\hspace{-0.08cm}+\hspace{-0.08cm}\frac{|h_{i,w}(k)|^2p_{i,w}(k)}{n_w(k)+\sum_{j:\mathbf{a}(j)=w, j\ne i}|h_{j,w}(k)|^2p_{j,w}(k)}\Big)\label{eqTransmissionRate}\\
&\hspace{-0.25cm}=\frac{1}{K}\hspace{-0.12cm}\sum_{k\in\mathcal{K}_w}\hspace{-0.12cm}
\log\hspace{-0.08cm}\Big(\hspace{-0.05cm}1+\frac{|h_{i,w}(k)|^2p_{i,w}(k)}{n_w(k)+I_{i}(k)}\Big)\triangleq
R_i(\mathbf{p}_{i,w},\mathbf{I}_{i,w};w)\label{eqRAlternativeDefinition}
\end{align}
where ${I}_i(k)$ denotes the aggregated received transmission power
level on channel $k$
 except CU $i$, i.e.,
\begin{align}
I_i(k)\triangleq\hspace{-0.5cm}\sum_{j:\mathbf{a}(j)=w, j\ne
i}\hspace{-0.4cm}|h_{j,w}(k)|^2p_{j,w}(k),~~
\mathbf{I}_{i,w}\triangleq\left\{I_i(k)\right\}_{k\in\mathcal{K}_w}\label{eqI}.
\end{align}
We note that, if $w=\mathbf{a}(i)$, then $\mathbf{I}_{i,w}$ can be
viewed as the set of interference {\it currently experienced} by CU
$i$; if $w\ne \mathbf{a}(i)$, $\mathbf{I}_{i,w}$ can  be viewed as
the set of interference that CU $i$ would experience {\it if it were
to switch to AP w}.

We see that \eqref{eqTransmissionRate} and
\eqref{eqRAlternativeDefinition} are equivalent definitions of the
CU $i$'s transmission rate. We will use either definition in the
following paragraph depending on the context.

\subsection{A Non-Cooperative Game Formulation}
We model each CU $i$ as selfish agent with the objective to find
strategy $(w^*,\mathbf{p}^{*}_{i,w^*})$ that maximizes its
transmission rate, based on the current state of the network:
\begin{align}
\hspace{-0.1cm}\big(w^*,\mathbf{p}^{*}_{i,w^*}\big)\in
\arg\max_{w\in\mathcal{W}}\max_{p_{i,w}\in\mathcal{F}_{i,w}}R_i(\mathbf{p}_{i,w},\mathbf{p}_{-i,w};
w).
\end{align}
We are now ready to define a non-cooperative game $\mathcal{G}$:
\begin{align}
\mathcal{G}\triangleq\left\{\mathcal{N},\{{\chi}_i\}_{i\in\mathcal{N}},
\{R_i\}_{i\in\mathcal{N}}\right\}\label{eqGame}
\end{align}
where the CUs $i\in\mathcal{N}$ are the players in the game; each
CU's strategy space can be expressed as $\chi_i\triangleq
\bigcup_{w\in\mathcal{W}}\left\{w, \mathcal{F}_{i,w}\right\}$;
each CU's utility function is its transmission rate
$R_i(\mathbf{p}_{i,w},\mathbf{p}_{-i,w}; w)$ as defined in
\eqref{eqTransmissionRate}. We emphasize that each feasible strategy
of a player contains a discrete variable and a continuous vector,
which makes the game $\mathcal{G}$ different from (and thus more
complicated than) most of the games considered in the context of
network resource allocation. We refer to the strategy space
$\{\chi_i\}_{i\in\mathcal{N}}$ of this game as {\it hybrid strategy
space}.

The NE of this game is defined as the tuple $\left\{\mathbf{a}^*(i),
\mathbf{p}^*_{i,{\mathbf{a}}^*(i)}\right\}_{i\in\mathcal{N}}$ such
that the following set of equations are satisfied:\vspace{-0.1cm}
\begin{align}
\hspace{-0.1cm}\Big(\mathbf{a}^*(i),\mathbf{p}^{*}_{i,\mathbf{a}^*(i)}\Big)\in
\arg\max_{w\in\mathcal{W}}\max_{p_{i,w}\in\mathcal{F}_{i,w}}R_i(\mathbf{p}_{i,w},\mathbf{p}^{*}_{-i,w};
w),~i\in\mathcal{N}\label{eqNE}
\end{align}
or equivalently,\vspace{-0.1cm}
$~\forall~i\in\mathcal{N},~w\in\mathcal{W},~\mathbf{p}_{i,w}\in\mathcal{F}_{i,w}$,
\begin{align}
R_i(\mathbf{p}_{i,\mathbf{a}^*(i)}^*,\mathbf{p}^{*}_{-i,\mathbf{a}^*(i)};
\mathbf{a}^*(i))\ge R_i(\mathbf{p}_{i,w},\mathbf{p}^{*}_{-i,w};
w)\nonumber.
\end{align}
Note that $\mathbf{p}^{*}_{-i,w}$ is defined as the power profiles
of all the CUs other than CU $i$ that is associated with AP $w$ in
the NE:
$\mathbf{p}^{*}_{-i,w}\triangleq\{\mathbf{p}^*_{j,\mathbf{a}^*(j)}\}_{j\ne
i, \mathbf{a}^*(j)=w}$. We call the equilibrium profile
$\mathbf{a}^*$ a {\it NE association profile}, and
$\mathbf{p}^*_{\mathbf{a}^*}\triangleq\big\{\mathbf{p}^*_{i,{\mathbf{a}}^*(i)}\big\}_{i\in\mathcal{N}}$
a {\it NE power allocation profile}. In order to avoid duplicated
definitions, we call the tuple $\left(\mathbf{a}^*,
\mathbf{p}^*_{\mathbf{a}^*}\right)$ a {\it joint equilibrium
profile} (JEP) of the game $\mathcal{G}$ (instead of a NE). It is
clear from either of the above definition that in a JEP, the system
is stable in the sense that no CU has the incentive to deviate from
either its AP association or its power allocation.

\section{Properties of the JEP}\label{secNE}
In this section, we introduce the notion of the potential function
for the game $\mathcal{G}$, and its relationship with the JEP. This
function plays an important role in our following analysis of the
existence of JEP and the proof of convergence of the algorithm. We
then prove that the JEP always exists for the game $\mathcal{G}$.

\subsection{The Potential Function}
Consider a simpler problem in which the association vector
$\mathbf{a}$ is {\it predetermined} and {\it fixed}. In this case,
the CUs do not need to choose their AP associations, thus the
problem of finding the JEP defined in \eqref{eqNE} reduces to the
one of finding the NE power allocation profile
$\mathbf{p}^*_{\mathbf{a}}$ that satisfies:
\begin{align}
\mathbf{p}^*_{i,\mathbf{a}(i)}\in\arg\max_{\mathbf{p}_i\in\mathcal{F}_{i,\mathbf{a}(i)}}R_i(\mathbf{p}_i,
\mathbf{p}^*_{-i,\mathbf{a}(i)};\mathbf{a}(i)),~~
\forall~i\in\mathcal{N}.\label{eqNEPowerProfile}
\end{align}
For a specific AP $w$, denote the set of CUs associated with it to
be $\mathcal{N}_w$: $\mathcal{N}_w\triangleq\{i:\mathbf{a}(i)=w\}$.
It is clear that $\{\mathcal{N}_w\}_{w\in\mathcal{W}}$ is a
partition of $\mathcal{N}$. We use
$\mathbf{p}_{w}\triangleq\left\{\mathbf{p}_{i,w}\right\}_{i\in\mathcal{N}_w}$
to denote the long vector containing the power profiles of all CUs
associated with AP $w$. When $\mathbf{a}$ is fixed, the activity of
the set of CUs $\mathcal{N}_w$, $w\in\mathcal{W}$ does not affect
the activity of the set of CUs $\mathcal{N}_q$, $q\in\mathcal{W},
q\ne w$, because of the fact that AP $w$ and $q$ operate on
different sets of channels. Consequently, the original game
$\mathcal{G}$ introduced in \eqref{eqGame} can be decomposed into
$W$ independent small games, with each small game
$\mathcal{G}^{\mathbf{a}}_w$ defined as:
\begin{align}
\mathcal{G}^{\mathbf{a}}_w=\left\{\mathcal{N}_w,\{{\mathcal{F}_{i,w}}\}_{i\in\mathcal{N}_w},
\{R_i\}_{i\in\mathcal{N}_w}\right\}.\label{eqSubGame}
\end{align}

Clearly, each of such small game has the same form as the spectrum
sharing game $G$ analyzed in Section III of the first part of the
paper. We define the potential function for the small game
$\mathcal{G}^{\mathbf{a}}_w$ as well as for the original game
$\mathcal{G}$ as follows.
\newtheorem{D1}{Definition}
\begin{D1}
{\it The {\it potential function of the game
$\mathcal{G}_w^{\mathbf{a}}$} under a feasible power profile
$\mathbf{p}_w$ is defined as:\vspace{-0.1cm}
\begin{align}
P_w(\mathbf{p}_w;{\mathbf{a}})=\frac{1}{K}\sum_{k\in\mathcal{K}_w}\left(\log\Big(n_w(k)+\sum_{i\in\mathcal{N}_w}|h_{i,w}(k)|^2
p_{i,w}(k)\Big)-\log n_w(k)\right)\nonumber.\vspace{-0.3cm}
\end{align}
The {\it system potential function} under a specific $\mathbf{a}$
and a feasible $\mathbf{p}$ is defined as the sum of the potential
functions associated to all games
$\{\mathcal{G}_w^{\mathbf{a}}\}_{w\in\mathcal{W}}$:
\begin{align}
P(\mathbf{p};\mathbf{a})=\sum_{w\in\mathcal{W}}P_w(\mathbf{p}_w;\mathbf{a}).
\end{align} }
\end{D1}

Define $\mathcal{F}^{\mathbf{a}}_{w}\triangleq
\prod_{i\in\mathcal{N}_w}\mathcal{F}_{i,w}$ as the joint feasible
set for the CUs that are associated with AP $w$ under the
association profile $\mathbf{a}$, and let
$\mathcal{F}^{\mathbf{a}}\triangleq
\prod_{w\in\mathcal{W}}\mathcal{F}^{\mathbf{a}}_{w}$. Let
$\mathcal{E}_w(\mathbf{a})$ denote the set of all NE power profiles
for the game $\mathcal{G}^{\mathbf{a}}_w$ \footnote{Indeed, as
argued in Section III and IV-A of \cite{hong10m_japsa_1}, the
spectrum sharing game $G$ (hence the small game
$\mathcal{G}^{\mathbf{a}}_w$) may have a connected set of NE power
profiles.}, then $\mathcal{E}(\mathbf{a})\triangleq
\prod_{w\in\mathcal{W}}\mathcal{E}_w(\mathbf{a})$ is the set of all
NE power profiles for the game $\mathcal{G}$ under fixed association
profile $\mathbf{a}$. Let $\mathbf{p}^*_w(\mathbf{a})$ be {\it any
one of } such NE power profiles for game
$\mathcal{G}_w^{\mathbf{a}}$, i.e.,
$\mathbf{p}^*_w(\mathbf{a})\in\mathcal{E}_w(\mathbf{a})$; let
$\mathbf{p}^*(\mathbf{a})$ be {\it any one of} the NE power profiles
of the network,
$\mathbf{p}^*(\mathbf{a})\in\mathcal{E}(\mathbf{a})$. The following
corollary regarding to the relationship between
$\mathbf{p}^*_w(\mathbf{a})$, $\mathbf{p}^*(\mathbf{a})$ and the
potential functions is a straightforward consequence of Theorem 1
and Corollary 1 of \cite{hong10m_japsa_1}.
\newtheorem{C1}{Corollary}
\begin{C1}\label{corollaryPotentialMaximization}
{\it For fixed $\mathbf{a}$, a feasible
$\mathbf{p}_{w}\in\mathcal{F}^\mathbf{a}_{w}$ maximizes the
potential function $P_w(\mathbf{p}_w;{\mathbf{a}})$ if and only if
it is in the set $\mathcal{E}_w(\mathbf{a})$.
We define the {\it maximum value} of the potential function:
\begin{align}
\bar{P}_w(\mathbf{a})\triangleq
\max_{\mathbf{p}_w\in\mathcal{F}^{\mathbf{a}}_w}P_w(\mathbf{p}_w;\mathbf{a})=
P_w(\mathbf{p}^*_w(\mathbf{a});\mathbf{a})\label{eqEP}
\end{align}
as  an equilibrium potential (EP) for AP $w$ under association
profile $\mathbf{a}$.

For a fixed $\mathbf{a}$, a feasible
$\mathbf{p}\in\mathcal{F}^{\mathbf{a}}$ that maximizes the system
potential function $P(\mathbf{p};\mathbf{a})$ if and only if it is
in the set $\mathcal{E}(\mathbf{a})$. Similarly as above, we refer
to the maximum value of the system potential function as the system
equilibrium potential (SEP) under association profile $\mathbf{a}$,
and denoted it by $\bar{P}(\mathbf{a})$:
\begin{align}
\bar{P}(\mathbf{a})\triangleq\sum_{w\in\mathcal{W}}\bar{P}_w(\mathbf{a}).\label{eqSEP}
\end{align}
}
\end{C1}


\subsection{Existence of JEP}
In this section, we discuss the existence of the JEP as defined in
\eqref{eqNE}. We emphasize here that determining the existence of
the JEP (which is a {\it pure} NE) for the game $\mathcal{G}$ is by
no means a trivial proposition. Due to the hybrid structure of the
game $\mathcal{G}$, the standard results on the existence of pure NE
of either continuous or discrete games can not be applied.
Consequently, we have to explore the
structure of the problem in proving the existence of JEP for the
game $\mathcal{G}$.

From Corollary \ref{corollaryPotentialMaximization} we see that a
specific $\mathbf{a}$ can be mapped to a SEP, denoted by
$\bar{P}(\mathbf{a})$. We claim that any one of the AP association
profiles $\widetilde{\mathbf{a}}$ that maximizes the SEP, along with
{\it any one} of its corresponding system power profile
$\mathbf{p}^*(\widetilde{\mathbf{a}})\in\mathcal{E}(\widetilde{\mathbf{a}})$,
constitute a JEP as defined in \eqref{eqNE}. We state this
observation in the following theorem.

\newtheorem{T1}{Theorem}
\begin{T1}\label{theoremExistence}
The game $\mathcal{G}$ as defined in \eqref{eqGame} always admits a
JEP. An association profile
$\widetilde{\mathbf{a}}\in\arg\max_{\mathbf{a}}\bar{P}(\mathbf{a})$,
along with any one of its corresponding NE power allocation profile
${\mathbf{p}}^*(\widetilde{\mathbf{a}})(j)=
\left\{\mathbf{p}^*_{i,\widetilde{\mathbf{a}}(i)}\right\}_{i\in\mathcal{N}}\in\mathcal{E}(\widetilde{\mathbf{a}})$,
constitute a JEP of the game $\mathcal{G}$ that satisfies
\eqref{eqNE}.
\end{T1}
\begin{proof}
We prove this theorem by contradiction. Suppose
$\widetilde{\mathbf{a}}$ maximizes the system potential, but
$\widetilde{\mathbf{a}}$ is not a NE association profile. Then there
must exist a CU $i$ who wants to switch from
$\widetilde{\mathbf{a}}(i)=\widetilde{w}$ to a different AP
$\widehat{w}\ne \widetilde{w}$. Define a new association profile
$\widehat{\mathbf{a}}$ as:
\begin{align}
\widehat{\mathbf{a}}(j)=\left\{ \begin{array}{ll}
\widetilde{\mathbf{a}}(j) &\textrm{for j}~\ne~i\\
\widehat{w} & \textrm{for j}~=i. \\
\end{array}\right.\label{eqAlphaContradiction}
\end{align}

Let
$\mathbf{p}^*(\widetilde{\mathbf{a}})\in\mathcal{E}(\widetilde{\mathbf{a}})$
and
$\mathbf{p}^*(\widehat{\mathbf{a}})\in\mathcal{E}(\widehat{\mathbf{a}})$.
The maximum rate that CU $i$ can get after switching to
$\widehat{w}$ {\it if all other CUs do not change their
actions}:{\small
\begin{align}
&\widehat{R}_i(\bar{\mathbf{p}}_{i,\widehat{w}},\mathbf{p}^*_{\widehat{w}}(\widetilde{\mathbf{a}});{\widehat{w}})\nonumber\\
&=\sum_{k\in\mathcal{K}_{\widehat{w}}}
\log\left(1+\frac{|h_{i,\widehat{w}}(k)|^2
\bar{p}_{i,\widehat{w}}(k)}{n_{\widehat{w}}(k)+\sum_{j:
\widetilde{\mathbf{a}}(j)=\widehat{w}}|h_{j,\widehat{w}}(k)|^2 p^*_{j,\widehat{w}}(k)}\right)\nonumber\\
&= \sum_{k\in\mathcal{K}_{\widehat{w}}}
\log\left(\frac{n_{\widehat{w}}(k)+I^*_{i}(k)+|h_{i,\widehat{w}}(k)|^2
\bar{p}_{i,\widehat{w}}(k)}{n_{\widehat{w}}(k)+I^*_{i}(k)}\right)\nonumber\\
&=P_{\widehat{w}}(\bar{\mathbf{p}}_{i,\widehat{w}},\mathbf{p}^*_{\widehat{w}}(\widetilde{\mathbf{a}});{\widehat{\mathbf{a}}})-
P_{\widehat{w}}(\mathbf{p}_{\widehat{w}}^*(\widetilde{\mathbf{a}});{\widetilde{\mathbf{a}}})\label{eqEstimatedRate}
\end{align}}
where $I^*_{i}(k)$ is defined similarly as in \eqref{eqI}, and the
vector $\bar{\mathbf{p}}_{i,\widehat{w}}$ is determined by:
\begin{align}
\bar{\mathbf{p}}_{i,\widehat{w}}=
\arg\max_{\mathbf{p}_i\in\mathcal{F}_{i,\widehat{w}}}
\widehat{R}_i(\mathbf{p}_i,\mathbf{p}^*_{\widehat{w}}(\widetilde{\mathbf{a}});{\widehat{w}}).
\end{align}
We can view the rate
$\widehat{R}_i(\bar{\mathbf{p}}_{i,\widehat{w}},\mathbf{p}^*_{\widehat{w}}(\widetilde{\mathbf{a}});{\widehat{w}})$
as CU $i$'s {\it estimate} of the maximum rate it can get if it were
to switch to AP $\widehat{w}$.

Because CU $i$ prefers $\widehat{w}$, from the definition of the JEP
\eqref{eqNE} we see that its current communication rate must be
strictly less than its estimated maximum rate, i.e., the following
must be true:
\begin{align}
{R}_i(\mathbf{p}^{*}_{i,\widetilde{w}}(\widetilde{\mathbf{a}}),\mathbf{p}^{*}_{-i,\widetilde{w}}(\widetilde{\mathbf{a}});{\widetilde{w}})<
\widehat{R}_i(\bar{\mathbf{p}}_{i,\widehat{w}},\mathbf{p}^*_{\widehat{w}}(\widetilde{\mathbf{a}});{\widehat{w}})
\label{eqRtLessThenRBarT+1}
\end{align}
where
$R_i(\mathbf{p}^{*}_{i,\widetilde{w}}(\widetilde{\mathbf{a}}),\mathbf{p}^{*}_{-i,\widetilde{w}}(\widetilde{\mathbf{a}});{\widetilde{w}})$
is the {\it actual} transmission rate for CU $i$ in the association
profile $\widetilde{\mathbf{a}}$, and it can be expressed as
follows:
\begin{align}
&{R}_i({\mathbf{p}}^*_{i,\widetilde{w}}(\widetilde{\mathbf{a}}),\mathbf{p}^*_{-i,\widetilde{w}}(\widetilde{\mathbf{a}});{\widetilde{w}})=\sum_{k\in\mathcal{K}_{\widetilde{w}}}
\log\left(1+\frac{|h_{i,\widetilde{w}}(k)|^2
{p}^*(k)}{n_{\widetilde{w}}(k)+I^*_{i}(k)}\right)\nonumber\\
&=P_{\widetilde{w}}({\mathbf{p}}^*_{\widetilde{w}}(\widetilde{\mathbf{a}});{\widetilde{\mathbf{a}}})-
P_{\widetilde{w}}(\mathbf{p}^*_{-i,\widetilde{w}}(\widetilde{\mathbf{a}});{\widetilde{\mathbf{a}}}).\label{eqEquilibriumRate}
\end{align}
Combining \eqref{eqEstimatedRate}, \eqref{eqRtLessThenRBarT+1} and
 \eqref{eqEquilibriumRate} we must have that:
\begin{align}
P_{\widetilde{w}}({\mathbf{p}}^*_{\widetilde{w}}(\widetilde{\mathbf{a}});{\widetilde{\mathbf{a}}})-&
P_{\widetilde{w}}(\mathbf{p}_{-i,\widetilde{w}}^*(\widetilde{\mathbf{a}});{\widetilde{\mathbf{a}}})<\nonumber\\
&P_{\widehat{w}}(\bar{\mathbf{p}}_{i,\widehat{w}},\mathbf{p}^*_{\widehat{w}}(\widetilde{\mathbf{a}});{\widehat{\mathbf{a}}})-
P_{\widehat{w}}(\mathbf{p}^*_{\widehat{w}}(\widetilde{\mathbf{a}});{\widetilde{\mathbf{a}}})\label{eqComparePotential1}.
\end{align}
We notice that the term
$P_{\widetilde{w}}(\mathbf{p}_{-i,\widetilde{w}}^*(\widetilde{\mathbf{a}});{\widetilde{\mathbf{a}}})$
is equivalent to
$P_{\widetilde{w}}(\mathbf{p}_{-i,\widetilde{w}}^*(\widetilde{\mathbf{a}});{\widehat{\mathbf{a}}})$
due to the equivalence of the following sets:
\begin{align}
\{j:j\ne i, \widetilde{\mathbf{a}}(j)=\widetilde{w}\}=\{j:j\ne i,
\widehat{\mathbf{a}}(j)=\widetilde{w}\}\label{eqNumberOfCUSameWHatWStar}.
\end{align}

Recall that from Corollary \ref{corollaryPotentialMaximization}, we
have that the NE power allocation profile maximizes the potential
function: $\mathbf{p}_{\widetilde{w}}^*(\widehat{\mathbf{a}})\in
\arg\max_{\mathbf{p}_{\widetilde{w}}\in\mathcal{F}^{\widehat{\mathbf{a}}}_{\widetilde{w}}}P_{\widetilde{w}}(\mathbf{p}_{\widetilde{w}};{\widehat{\mathbf{a}}})$.
Observe that the set of CUs associated with AP $\widetilde{w}$ under
profile $\widehat{\mathbf{a}}$ is the same as the set of CUs
associated with AP $\widetilde{\mathbf{a}}$ under profile
$\widetilde{\mathbf{a}}$ excluding CU $i$, we must have
$\mathbf{p}_{-i,\widetilde{w}}^*(\widetilde{\mathbf{a}})\in\mathcal{F}^{\widehat{\mathbf{a}}}_{\widetilde{w}}$.
Consequently, the following is true:
\begin{align}
 P_{\widetilde{w}}({\mathbf{p}}^*_{\widetilde{w}}(\widehat{\mathbf{a}});{\widehat{\mathbf{a}}})& \ge
 P_{\widetilde{w}}(\mathbf{p}^*_{-i,\widetilde{w}}(\widetilde{\mathbf{a}});{\widehat{\mathbf{a}}})
 \stackrel{(a)}=P_{\widetilde{w}}(\mathbf{p}_{-i,\widetilde{w}}^*(\widetilde{\mathbf{a}});\widetilde{\mathbf{a}})\label{eqPotentialOPT}
\end{align}
 where $(a)$ is from \eqref{eqNumberOfCUSameWHatWStar}. Similarly, we have
 that:
\begin{align}
P_{\widehat{w}}({\mathbf{p}}^*_{\widehat{w}}(\widehat{\mathbf{a}});{\widehat{\mathbf{a}}})\ge
P_{\widehat{w}}(\bar{\mathbf{p}}_{i,\widehat{w}},\mathbf{p}_{\widehat{w}}^*(\widetilde{\mathbf{a}});{\widehat{\mathbf{a}}})
.\label{eqPotentialOPT+1}
\end{align}
Combining \eqref{eqPotentialOPT}, \eqref{eqPotentialOPT+1} and
\eqref{eqComparePotential1}, we have that:{\small
\begin{align}
P_{\widetilde{w}}({\mathbf{p}}_{\widetilde{w}}^*(\widetilde{\mathbf{a}});{\widetilde{\mathbf{a}}})-
P_{\widetilde{w}}({\mathbf{p}}^*_{\widetilde{w}}(\widehat{\mathbf{a}});{\widehat{\mathbf{a}}})<
P_{\widehat{w}}({\mathbf{p}}^*_{\widehat{w}}(\widehat{\mathbf{a}});{\widehat{\mathbf{a}}})-
P_{\widehat{w}}(\mathbf{p}_{\widehat{w}}^*(\widetilde{\mathbf{a}});{\widetilde{\mathbf{a}}})\nonumber
\end{align}}
which essentially says that after $i$ switched to AP $\widehat{w}$,
the decrease of EP of AP $\widetilde{w}$ is less than the increase
of the EP of AP $w$. In other words, we have that:{\small
\begin{align}
P_{\widetilde{w}}({\mathbf{p}}_{\widetilde{w}}^*(\widetilde{\mathbf{a}});\widetilde{\mathbf{a}})+
P_{\widehat{w}}(\mathbf{p}_{\widehat{w}}^*({\widetilde{\mathbf{a}}});
\widetilde{\mathbf{a}})<
P_{\widehat{w}}({\mathbf{p}}^*_{\widehat{w}}(\widehat{\mathbf{a}});{\widehat{\mathbf{a}}})+
P_{\widetilde{w}}({\mathbf{p}}^*_{\widetilde{w}}(\widehat{\mathbf{a}});{\widehat{\mathbf{a}}}).
\label{eqTwoTermSumPotential}
\end{align}}
Noticing that the equilibrium potentials of all the APs other than
$\widetilde{w}$ and $\widehat{w}$ are the same between the profile
$\widetilde{\mathbf{a}}$ and $\widehat{\mathbf{a}}$, thus adding
them to both sides of \eqref{eqTwoTermSumPotential} we have that:
\begin{align}
\sum_{w\in\mathcal{W}}P_{w}({\mathbf{p}}^*_{
{w}}(\widetilde{\mathbf{a}});{\widetilde{\mathbf{a}}})<
\sum_{w\in\mathcal{W}}P_w({\mathbf{p}}^*_w(\widehat{\mathbf{a}});{\widehat{\mathbf{a}}})\label{eqSumPotentialInequality}
\end{align}
which is equivalent to:
\begin{align}
\bar{P}(\widetilde{\mathbf{a}})<\bar{P}(\widehat{\mathbf{a}})\label{eqContradiction}.
\end{align}
This is a contradiction to the assumption that
$\bar{P}(\widetilde{\mathbf{a}})$ maximizes the system potential. We
conclude that $\widetilde{\mathbf{a}}$ must be a NE association
profile. Clearly, $\mathbf{p}^*(\widetilde{a})$ is a NE power
allocation profile. Consequently, we have that
$\left(\widetilde{\mathbf{a}}, \mathbf{p}^*(\widetilde{\mathbf{a}})
\right)$ is a JEP.
\end{proof}

\section{The Proposed Algorithm}\label{secAlgorithm}
In this section, we introduce our main algorithm, referred to as the
Joint Access point Selection and Power Allocation (JASPA) algorithm,
that allows the CUs in the network to distributely compute the JEP.
To this end, we first introduce a simple scheme that assigns the CUs
to their closest AP, a scheme which essentially separates the
process of AP association and power allocation. This scheme,
although relatively simple, offers valuable insights upon which we
build the JASPA algorithm, in subsection \ref{subJASPA}.

\subsection{Closest AP Association Algorithms}\label{subNearestAP}
Consider a fixed AP association profile $\mathbf{a}$ in which each
CU is assigned to its closest AP. The ``closeness", or ``distance"
from a CU to the APs can be measured either by the physical distance
between them, or by the strength of pilot/control signal received by
the CU from the AP. Assuming that each CU has a single closest AP
\footnote{This assumption is without loss of generality because if
two APs have the same ``distance" to a CU, they can be further
ranked by other closeness criterion.}, then the AP association
profile is unique and the computation of JEP reduces to the problem
of finding the NE power allocation profile. Moreover, as mentioned
before, the CUs are partitioned into independent sets
$\mathcal{N}_w\triangleq\{i:\mathbf{a}(i)=w\}$, and the CUs in each
set $\mathcal{N}_w$ can compute their NE power allocation profile
without taking into consideration the behaviors of the CUs in other
sets.

Clearly, this scheme {\it separates} the process of spectrum
decision and spectrum sharing, and the CUs only need to carry out
the task of sharing the spectrum available to the designated AP with
other CUs. However, as we probably can speculate, no matter how
efficient such sharing scheme is, the overall system performance
might suffer because of the fixed and inefficient AP assignment. We
will see such performance degradation later in the simulation
section.


We note that from Proposition 2 and 3 in \cite{hong10m_japsa_1}, for
a specific association profile $\mathbf{a}$, the set of CUs
$\mathcal{N}_w$ that is associates to the same AP $w$ are able to
distributedly decide on their NE power allocation profiles by
running either the A-IWF or the S-IWF algorithm (cf. Algorithm 1 and
Algorithm 2 in \cite{hong10m_japsa_1}).
%

\subsection{The Joint AP Selection and Power Allocation
Algorithm}\label{subJASPA}
 We name the proposed algorithm Joint
Access Point Selection and Power Allocation (JASPA) algorithm.
Intuitively, the proposed algorithm works as follows. For a fixed AP
association profile, all CUs calculate iteratively their NE power
allocations. After convergence, they individually try to see if they
can {\it strictly} increase their communication rates by switching
to another AP, {\it assuming that all other CUs keep their current
AP associations and power profiles}. When CU $i$ decides that its
next best AP association should be $w_i^*$, we record his decision
by a $W\times 1$ {\it best reply vector}
$\mathbf{b}_i:\mathbf{b}_i=\mathbf{e}_{w^*_i}$, where
$\mathbf{e}_{j}$ denotes a $W\times 1$ elementary vector with all
entries $0$ except for the $j^{th}$ entry, which takes the value
$1$. In the next iteration, CU $i$'s actual AP association decision
is made according to a $W\times 1$ {\it probability vector}
$\bfbeta_i^t$, which is properly updated in each iteration according
to $\mathbf{b}_i$. We also suppose that each CU has a length $M$
memory, operated in a first in first out (FIFO) fashion, that
records its last $M$ best reply vectors.

The proposed algorithm is detailed as follows.

1) {\bf Initialization}: Let t=0, CUs randomly choose their APs.

2) {\bf Calculation of the NE Power Allocation Profile}: Based on
the current association $\mathbf{a}^t$, all the CUs calculate their
NE power allocations $\mathbf{p}^{*}_{i}(\mathbf{a}^t)$, either by
A-IWF or S-IWF algorithm. We call the process of reaching such
intermediate equilibrium an ``inner loop".

3) {\bf Selection of the Best AP Association}: Each CU $i$ talks to
all the APs in the network, obtains necessary information in order
to find a set of APs $\mathcal{W}^t_i$ such that all
$w\in\mathcal{W}^t_i$ satisfies $w\ne\mathbf{a}^t(i)$ and:{
\begin{align}
\hspace{-0.5cm}\max_{\mathbf{p}_{i,w}\in\mathcal{F}_{i,w}}R_i(\mathbf{p}_{i,w},\mathbf{p}^*_{w}(\mathbf{a}^t);w)>
R_i(\mathbf{p}^*_{i}(\mathbf{a}^t),\mathbf{p}^*_{-i}(\mathbf{a}^t);{\mathbf{a}^t(i)}).\label{eqBetterAP}
\vspace{-0.5cm}
\end{align}
}
If $\mathcal{W}^t_i\ne\emptyset$, obtain the
$w^*_i\in\mathcal{W}^t_i$ that can offer the maximum rate (ties are
randomly broken); otherwise, let $w^*_i=\mathbf{a}^t(i)$. Set the
best reply vector $\mathbf{b}^{t+1}_i=\mathbf{e}_{w^*_i}$.

4) {\bf Update Probability Vector}: For each CU $i$, update the
$W\times 1$ probability vector $\bfbeta^t_i$ according to:
\begin{align}
\bfbeta^{t+1}_i=\left\{ \begin{array}{ll} \bfbeta
^t_i+\frac{1}{M}(\mathbf{b}^{t+1}_i-
\mathbf{b}^{t-M}_i)&~\textrm{if}~M \le t\\
\bfbeta ^t_i+\frac{1}{M}(\mathbf{b}^{t+1}_i-
\mathbf{b}^{1}_i)&~\textrm{if}~M>t>0\\
\mathbf{b}^{1}_i &\textrm{if}~t=0. \\
\end{array} \right.\label{eqUpdateBeta}
\end{align}
Shift $\mathbf{b}_i^{t+1}$ into the end of the memory; shift
$\mathbf{b}^{t-M}_i$ out from the front of the memory if $t\ge M$.

5) {\bf Determine the Next AP Association}: Each CU $i$ samples the
AP index for association at next iteration according to the
probability $\bfbeta^{t+1}_i$, i.e.,
\begin{align}
\mathbf{a}^{t+1}(i)\sim multi(\bfbeta^{t+1}_i)\label{eqSampleA}
\end{align}
where $multi(.)$ represents a multinomial distribution.

6) {\bf Continue}: Let t=t+1, and go to Step 2).

%

We make several comments regarding to the above JASPA algorithm.

\newtheorem{R1}{Remark}
\begin{R1} It is crucial that each CU finally decides on choosing a
{\it single} AP for transmission. Failing to do so will result in
system instability, in which the CUs switch AP association
indefinitely, and much of the system resource will be wasted for
closing old connections and re-establishing new connections between
the APs and CUs. In another word, it is preferable that for all
$i\in\mathcal{N}$, $ \lim_{t\to\infty}\bfbeta^t_i=\bfbeta^*_i $ and:
\begin{align}
\bfbeta^*_i(w)=\left\{ \begin{array}{ll}
1 &\textrm{for a single}~w\in\mathcal{W}\\
0 & \textrm{otherwise.}
\end{array}\right.
\end{align}
\end{R1}

\newtheorem{R2}{Remark}
\begin{R1}
The best reply vectors $\{\mathbf{b}^{t+1}_i\}_{i\in\mathcal{N}}$
are decided in each iteration based on the other CUs' AP
associations and power profiles in the previous iteration.  It is
straightforward to show that in order to calculate
$\max_{\mathbf{p}_{i,w}\in\mathcal{F}_{i,w}}R_i(\mathbf{p}_{i,w},\mathbf{p}^*_{w}(\mathbf{a}^t);w)$
for different $w\in\mathcal{W}$, individual CU $i$ does not need to
know the strategies of all other CUs in the network, nor does it
need to know the system association profile $\mathbf{a}^t$. Instead,
it only requires the information of {\it aggregated interference
plus noise} on each channel from each AP of the last iteration. This
is precisely the necessary information needed for finding the set
$\mathcal{W}^t_i$ in Step 3) of the JASPA. This property of the
algorithm contributes to the reduction of the amount of messages
exchanged between APs and each CU when making association decisions.
\end{R1}
\newtheorem{R3}{Remark}
\begin{R1}
Considering the overhead regarding to end an old connection and
re-establish a new connection, it is reasonable to assume that a
selfish CU is unwilling to abandon its current AP if the new one
cannot offer {\it significant} improvement of the data rate. We can
model such unwillingness of the CUs by introducing a {\it connection
cost} $c_i\ge 0$, which is a private parameter for each CU $i$. A CU
$i$ will only seek to switch to a new AP if the new one can offer
rate improvement of at least $c_i$, i.e., it will only switch to
those APs $w\in\mathcal{W}^t_i$ that satisfies:{
\begin{align}
\max_{\mathbf{p}_{i,w}\in\mathcal{F}_{i,w}}R_i(\mathbf{p}_{i,w},\mathbf{p}^*_{w}(\mathbf{a}^t);w)\ge
R_i(\mathbf{p}^*_{i}(\mathbf{a}^t),\mathbf{p}^*_{-i}(\mathbf{a}^t);{\mathbf{a}^t(i)})+c_i.\nonumber
\end{align}}
From a system point of view, such unwillingness to switch by the CUs
might contribute to improved convergence speed of the algorithm, but
might also result in reduced system throughput. These two
phenomenons are indeed observed in our simulations, please see
section \ref{secSimluation} for examples. We note that the the
equilibrium solution resulted from using the costs
$\{c_i\}_{i\in\mathcal{N}}$ is closely related to the notion of
``$\epsilon$-equilibrium" in the game theory. See chapter 4 of
\cite{basar99} for details.
\end{R1}

\subsection{Proof of Convergence}\label{subsecConvergence} In this
section, we prove that the JASPA algorithm converges to a JEP {\it
globally}, i.e., the algorithm converges regardless of the initial
starting points of the algorithm, or the realizations of the channel
gains.

We first introduce some notations. Let $\mathbf{c}^t$ be a vector
denoting the {\it best reply association profile} at time $t$, i.e.,
$\mathbf{c}^t(i)=w$ if and only if $\mathbf{b}_i^{t}(w)=1$. Define a
set $\mathcal{C}$ and $\mathcal{A}$ as follows: $ \mathbf{c}\in
\mathcal{C} \Longleftrightarrow \mathbf{c} \textrm{~~infinitely
often in ~} \{\mathbf{c}^t\}_{t=1}^{\infty}. $ and $ \mathbf{a}\in
\mathcal{A} \Longleftrightarrow \mathbf{a} \textrm{~~infinitely
often in ~} \{\mathbf{a}^t\}_{t=1}^{\infty}$. We first provide a
proposition stating that there must exist a NE association profile
$\mathbf{a}^{*}$ that satisfies $\mathbf{a}^*\in\mathcal{A}$. The
proof of this proposition can be found in Appendix \ref{app1}.

\newtheorem{P1}{Proposition}
\begin{P1}\label{propIO}
{\it Choose $M\ge N$. Then at least one element in the set
$\mathcal{A}$, say $\mathbf{a}^{*}$, is a NE association profile.
Moreover, $(\mathbf{a}^{*},\mathbf{p}^{*}(\mathbf{a}^*))$ is a JEP
(satisfy equation \eqref{eqNE}).}
\end{P1}

Using the result in Proposition \ref{propIO}, we obtain the
following convergence results.

\newtheorem{T5}{Theorem}
\begin{T1}\label{theoremConvergenceJASPA}
{\it When choosing $M\ge N$, the JASPA algorithm produces a sequence
$\left\{(\mathbf{a}^t,
\mathbf{p}^*(\mathbf{a}^t))\right\}_{t=1}^{\infty}$ that converges
to a JEP $(\mathbf{a}^*,\mathbf{p}^*(\mathbf{a}^*))$ with
probability 1.}
\end{T1}
\begin{proof}
We first show that the sequence
$\left\{\mathbf{a}^t\right\}_{t=1}^{\infty}$ converges to an
equilibrium profile $\mathbf{a}^*$. Notice that if at time $T$,
$\mathbf{a}^{T}=\mathbf{a}^*$, and in the next $M$ iterations, we
always have $\mathbf{a}^{T+t}=\mathbf{a}^*,~t=1,\cdots,M$, then the
algorithm converges.

Let $\mathcal{A}^*\in\mathcal{A}$ contains all the NE association
profiles in $\mathcal{A}$. Let $\{\mathbf{a}^{t(k)}:k\ge 1\}$ be the
infinite subsequence satisfying $\mathbf{a}^{t(k)}\in\mathcal{A}^*$.
Without loss of generality, assume $t(k)-t(k-1)\ge M$. Let us denote
by $C_k$ the event in which the process converges to a
$\mathbf{a}^*\in\mathcal{A}^*$, after a sequence of best replies
equals to $\mathbf{a}^*$ of length $M$ occurs, starting at time
$t(k)$: $ C_k=\bigcap_{l=1}^{M}\{\mathbf{a}^{t(k)+l}=\mathbf{a}^*\}.
$ Note, $ Pr(C_{k+1}|C_k^c)\ge(\frac{1}{M})^{N\times M}$, because
whenever $\mathbf{a}^*$ appears, each CU $i$'s best reply should be
$\mathbf{a}^*(i)$, hence $\mathbf{a}^*(i)$ will be inserted into the
last slot of CU $i$'s memory. Then with probability
$(\frac{1}{M})^N$, all CUs sample the last memory and $\mathbf{a}^*$
will appear in the next iteration. Thus,{\small
\begin{align}
\hspace{-0.1cm}&Pr\hspace{-0.1cm}\left(\bigcap_{k\ge
1}C^c_k\right)\hspace{-0.15cm}=\hspace{-0.15cm}\lim_{T\to\infty}\hspace{-0.15cm}Pr\hspace{-0.1cm}\left(\bigcap_{k=1}^{
T}C^c_k\right)\hspace{-0.1cm}=\hspace{-0.1cm}\lim_{T\to\infty}\prod_{k=1}^{T-1}\hspace{-0.15cm}\left(1-Pr(C_{k+1}|C^c_k)\right)\nonumber\\
\hspace{-0.1cm}&\le\lim_{T\to\infty}\Big(1-(\frac{1}{M})^{n\times
M}\Big)^{T-1}\hspace{-0.3cm}=0.
\end{align}}
This says $ Pr(\mathbf{a}^t ~\textrm{converges to a }
\mathbf{a}^*\in\mathcal{A}^*~\textrm{eventually})=1. $ Finally,
because $\mathbf{p}^*(\mathbf{a}^*)\in\mathcal{E}(\mathbf{a}^*)$ is
a NE power allocation profile, we conclude that
$\left(\mathbf{a}^*,\mathbf{p}^*(\mathbf{a}^*)\right)$ is a JEP.
\end{proof}

We mention that the requirement on the length of the memory is
technical in order to facilitate the proof. In simulations, we
observe that such requirement is not necessary for ensuring
convergence.

Now that we have shown the convergence of the JASAP to the JEP, it
is of interest to evaluate the ``quality" of such network
equilibrium. In this work, we use the system throughput to measure
the quality of the JEP, and our simulation results (to be shown in
section \ref{secSimluation}) are very encouraging.


\section{Extensions to the JASPA Algorithm}\label{secExtension}

The JASPA algorithm presented in the previous section is
``distributed" in the sense that the computation that each CU needs
to carry out in each iteration only requires some local/summary
information, i.e., the aggregated interference plus noise at
different APs in different channels, and the CU's own channel gain.
However, this algorithm requires that for each AP association
profile $\mathbf{a}^t$, an {\it intermediate} equilibrium
$\mathbf{p}^*(\mathbf{a}^t)$ should be reached, and at each
iteration $t$ the CUs {\it cannot} choose their next AP association
profile {\it until} the system reaches such equilibrium. This
requirement poses a relatively strong level of coordination among
the CUs (although this issue can be alleviated by letting the APs
orchestrate the updating instances), which is not entirely desirable
for a distributed algorithm.

In this section, we propose two algorithms that do not require that
the CUs reach any intermediate equilibria. Specifically, we propose
1) a sequential version of the JASPA algorithm (Se-JASPA) in which
CUs act one by one in each step, and 2) a simultaneous/parallel
version of the JASPA algorithm (Si-JASPA) in which CUs act at the
same time.

The Se-JASPA algorithm is detailed in Table \ref{tableSeJASPA}.
\begin{table}[ht]
\begin{center}
\vspace{-0.4cm}
\begin{tabular*}{ 0.5\textwidth }{l}
\hline\\
1) {\bf Initialization (t=0)}: Each CU randomly chooses
$\mathbf{a}^0(i)$ and
$\mathbf{p}_{i,\mathbf{a}^0(i)}^0$\\
2) {\bf Determine the Next AP Association}:\\
~~If it is CU $i$'s turn to act, (e.g., $\{(t+1) \textrm{mode} N
\}+1=i $), then CU $i$ \\
~~finds a set $ \mathcal{W}^*_i$ s.t.:\\
~~~~~~~~$ \mathcal{W}^*_i=
\arg\max_{w\in\mathcal{W}}\max_{\mathbf{p}_{i,w}\in\mathcal{F}_{i,w}}
R(\mathbf{p}_{i,w},\mathbf{p}^t_{w};w)\label{eqUpdateASequential}
$\\
~~Then it selects an AP by randomly picking
$w^*\in\mathcal{W}^*_i$ and setting $\mathbf{a}^{t+1}(i)=w^*$.\\
~~For other CUs $j\ne i$, $\mathbf{a}^{t+1}(j)=\mathbf{a}^{t}(j)$\\
3) {\bf Update the Power Allocation}: \\
~~Denote $w^*=\mathbf{a}^{t+1}(i)$, Then CU $i$ calculates $\mathbf{p}^{t+1}_i$ as\\
~~~~$ \mathbf{p}^{t+1}_i=\left\{ \begin{array}{l} \arg
\max_{\mathbf{p}_{i,w^*}\in\mathcal{F}_{i,w^*}}R_i(\mathbf{p}_{i,w^*},\mathbf{p}^t_{w^*};{w^*}),~
\textrm{if}~ w^*\ne \mathbf{a}^{t}(i) \\
\arg
\max_{\mathbf{p}_{i,w^*}\in\mathcal{F}_{i,w^*}}R_i(\mathbf{p}_{i,w^*},\mathbf{p}^t_{-i,w^*};{w^*}),
~\textrm{otherwise}\\
\end{array} \right.\label{eqUpdatePSequential}
$ \\
~~For other CUs $j\ne i$, $\mathbf{p}_j^{t+1}=\mathbf{p}_j^t$\\

4) {\bf Continue}: Let t=t+1, and go to Step 2)\\

 \hline
\end{tabular*}
\caption{The Se-JASPA Algorithm} \label{tableSeJASPA}
\end{center}
\vspace{-0.9cm}
\end{table}

We partially characterize the convergence behavior of Se-JASPA
algorithm in the following theorem, the proof of which can be found
in Appendix \ref{app2}.

\newtheorem{T6}{Theorem}
\begin{T1}\label{theoremConvergenceSeJASPA}
{\it The sequence of system potential
$\{P(\mathbf{p}^t,\mathbf{a}^t)\}^{\infty}_{t=1}$ produced by the
Se-JASPA algorithm is non-decreasing and converging.}
\end{T1}

Some brief comments regarding to the Se-JASPA algorithm is in order.
We see that the Se-JASPA algorithm differs from the JASPA algorithm
in several important ways. Firstly, a CU $i$ does not need to keep
its best reply vector $\mathbf{b}^t$ as it does in JASPA. It decides
on its AP association greedily in step 2). Secondly, a CU $i$, after
deciding a new AP $\mathbf{a}^{t+1}(i)=w^*$, does not need to go
through the process of reaching an intermediate equilibrium with all
other CUs to obtain $\mathbf{p}^{t+1}_{i,w^*}$. However, the CUs
still need to be coordinated for the exact sequence of their update,
because in each iteration only a single CU is allowed to act. Such
order of update can be agreed upon and enforced by the APs in the
network. As might be inferred by the sequential nature of this
algorithm, when the number of CUs is large, the convergence becomes
slow.

The Si-JASPA algorithm, as detailed in Table \ref{tableSiJASPA},
overcomes the above difficulties encountered in Se-JASPA. We note
that in the algorithm, the variable $T_i$ represents the duration
that CU $i$ has stayed in the current AP, and the stepsizes
$\{\alpha_t\}$ is similarly defined as in the A-IWF algorithm in
\cite{hong10m_japsa_1}: $\alpha_t\in(0,1)$ and
\begin{align}
\lim_{T\to\infty}\sum_{t=1}^{T}\alpha_t=\infty,
~\lim_{T\to\infty}\sum_{t=1}^{T}\alpha^2_t<\infty.\label{eqAlphaProperty}
\end{align}

\begin{table}
\begin{center}
\vspace{-0.1cm}
\begin{tabular*}{ 0.5\textwidth }{l}
\hline\\
1) {\bf Initialization (t=0)}: Each CU $i$ randomly chooses
$\mathbf{a}^0(i)$ and $\mathbf{p}_{i,\mathbf{a}^0(i)}^0$
\\
2) {\bf Selection of the Best Reply Association}:\\
~~Each CU obtains the AP $w^*_i$ and set $\mathbf{b}^{t+1}_i$ following Step 3) of JASPA\\
3) {\bf Update Probability Vector}: \\
~~Each CU $i$ updates the probability vector $\bfbeta^t_i$ according
to \eqref{eqUpdateBeta} \\
~~Shift $\mathbf{b}_i^{t+1}$ into the memory; shift
$\mathbf{b}^{t-M}_i$ out of memory if $t\ge M$
\\
4) {\bf Determine the Next AP Association}: \\
~~Each CU $i$ samples the AP index for association as in
\eqref{eqSampleA}
\\
5) {\bf Compute the Best Reply Power Allocation}: \\
~~Let $w^{t+1}_i=\mathbf{a}^{t+1}(i)$. Each CUs $i$ calculates
$\mathbf{p}^*_i$ as\\
~~~~~$
\mathbf{p}^*_i=\max_{\mathbf{p}_{i,w^{t+1}_i}}R_i(\mathbf{p}_{i,w^{t+1}_i},\mathbf{p}^t_{-i,w^{t+1}_i};{w^{t+1}_i})
$
\\
6) {\bf Update the Duration of Stay}: \\
~~Each CU $i$ maintains and updates a variable $T_i$: \\
~~~~${T}_i=\left\{
\begin{array}{ll}
1&\textrm{if}~\mathbf{a}^{t+1}(i)\ne\mathbf{a}^{t}(i)\\
T_i+1&\textrm{if}~\mathbf{a}^{t+1}(i)=\mathbf{a}^{t}(i)\\
\end{array} \right.
$
\\
7) {\bf Update the Power Allocation}: \\
~~Each CU $i$ calculates $\mathbf{p}^{t+1}_i$ as follows:
\\
~~~~$\mathbf{p}^{t+1}_i=\left\{
\begin{array}{ll}
\mathbf{p}^*_i&~\textrm{if}~\mathbf{a}^{t+1}(i)\ne\mathbf{a}^{t}(i)\\
(1-\alpha_{T_i})\mathbf{p}^t_i+\alpha_{T_i}\mathbf{p}^*_i&~~\textrm{if}~\mathbf{a}^{t+1}(i)=\mathbf{a}^{t}(i)\\
\end{array} \right.\label{eqUpdatePSimultaneous}
$\\
8) {\bf Continue}: Let t=t+1, and go to Step 2)\\
 \hline
\end{tabular*}
\caption{The Si-JASPA Algorithm} \label{tableSiJASPA}
\end{center}
\vspace{-2cm}
\end{table}

We see that the structure of the Si-JASPA is almost the same as the
JASPA except that each CU, after switching to a new AP, does not
need to go through the process of joint computation of the
intermediate equilibrium solution. Instead, the CUs can make their
AP decision ``continuously". The level of coordination among the CUs
required for this algorithm is minimum among all the three
algorithms introduced so far. The simultaneous update required by
this algorithm can be realized by either one of the following
approaches:
\begin{itemize}
\item The APs agree upon the update interval off-line. Each CU is equipped with a timer.
The first time a CU comes into the system, it is informed by its
initial associated AP the update interval and the next update
instance. After that, this CU can perform update on its own.
\item The APs agree upon the update interval off-line. When the time
comes for the update, the APs individually alert the CUs associated
with them by broadcasting.
\end{itemize}

Extensive simulations suggest that this algorithm converges faster
than the Se-JASPA.

\section{JASPA Based on Network-Wide Joint-Strategy }
\label{secJJASPA}

The Se/Si-JASPA algorithms introduced in the previous section
reliefs the CUs from the burden of reaching intermediate
equilibrium. However, the lack of general proof of convergence for
them might be a concern to us (although they appear to be always
convergent in practice). In this section, an alternative algorithm
with convergence guarantee is proposed. This algorithm allows the
CUs, as in the Se/Si-JSPA, to jointly select their power profiles
and AP association without the need to reach the intermediate
equilibria. We will see later that compared with all the algorithms
introduced previously, the algorithm studied in this section
requires considerably different information/memory structure for
both the CUs and the APs. Among others, it requires that the CUs
maintain in their memory some history of the network-wide {\it joint
strategy} of all CUs. We henceforth name this algorithm
Joint-strategy JASPA (J-JASPA).

\subsection{The J-JASPA Algorithm}
We first give some definitions. As all previously mentioned
algorithms, the J-JASPA algorithm is iterative in nature, thus in
the following we use $t$ to denote the $t^{th}$ iteration of the
algorithm, if needed.
\begin{itemize}
\item Let
$\mathcal{N}^t_w\triangleq\left\{i:\mathbf{a}^t(i)=w\right\}$ be the
set of CUs that are associated with AP $w$ in iteration $t$.

\item Let $\mathcal{Q}$ be any subset
of $\mathcal{N}$. Define the {\it last time} that the subset
$\mathcal{Q}$ of CUs is associated with a particular AP $w$ as
$\widehat{t_w}(\mathcal{Q})$, i.e.,
\begin{align}
\widehat{t_w}(\mathcal{Q})=\left\{
\begin{array}{ll}
\arg\max_{t}\{\mathcal{N}^t_w=\mathcal{Q}\}&\textrm{if}~
\bigcup_{t\ge 0}\left\{\mathcal{N}^t_w=\mathcal{Q}\right\}\ne\emptyset\\
\infty&\textrm{otherwise.}\\
\end{array} \right.
\end{align}

\item Let $\mathbf{I}(\mathcal{N}_w)$ be the joint interference profile
by the subset of CUs $\mathcal{N}_w$ that is associated with AP $w$:
$\mathbf{I}(\mathcal{N}_w)\triangleq
\left\{\mathbf{I}_{i,w}\right\}_{i\in\mathcal{N}_w}$, where
$\mathbf{I}_{i,w}$ is defined in \eqref{eqI}.
\end{itemize}

As we mentioned before, one of the distinct feature of J-JASPA
algorithm is the information/memory structure required for carrying
out the computation. Specifically, each CU $i$ keeps three different
memories, each of which is of length $M$ and operates in a FIFO
fashion. The first memory, referred to as {\it association memory}
(AM), records CU $i$'s last $M$ associated AP
$\left\{\mathbf{a}^t(i)\right\}^{T}_{t=T-M+1}$, i.e.,
$AM_i(m)=\mathbf{a}^{T-M+m}(i)$. Here we use $AM_i(m)$ to denote
$m^{th}$ element in CU $i$'s AM. The second memory, referred to as
{\it interference memory} (IM), records the last $M$ system
interference levels for CU $i$,
$\left\{\mathbf{I}_i^t\right\}^{T}_{t=T-M+1}$, where
$\mathbf{I}^t_i\triangleq\{\mathbf{I}^t_{i,w}\}_{w\in\mathcal{W}}$.
The third memory, referred to as {\it rate memory}, records the last
$M$ CU $i$'s sum rate,
$\left\{R_i\left(\mathbf{p}^t_{\mathbf{a}^t(i)};
\mathbf{a}^t(i)\right)\right\}^{T}_{t=T-M+1}$.

Each AP $w$ is also required to keep track of some local quantities
\footnote{Here, we use ``local" to signify the fact that individual
AP can gather these information without the need to communicate with
other APs.} regarding to the history of the CU behaviors.
Specifically, AP $w$ keeps track of the following variables for each
subset $\mathcal{Q}\subseteq\mathcal{N}$ that has been associated
with AP $w$ during time $[0,~~T]$ at least once:
\begin{itemize}
\item The local power  profile $\mathbf{p}(\mathcal{Q})=
\left\{\mathbf{p}^{\widehat{t_w}(\mathcal{Q})}_{i}\right\}_{i\in\mathcal{N}_w}$.
\item The local interference profile $\mathbf{I}(\mathcal{Q})=
\left\{\mathbf{I}^{\widehat{t_w}(\mathcal{Q})}_{i,w}\right\}_{i\in\mathcal{N}_w}$.
\item The total number of times that $\mathcal{Q}$ has been played: $\widehat{T}(\mathcal{Q})=\sum_{t\le
T}\mathbf{1}\{\mathcal{N}^t_w=\mathcal{Q}\}$, where
$\mathbf{1}\{.\}$ is the indicator function.
\end{itemize}

Then the J-JASPA algorithm can be detailed as follows: \\
1) {\bf Initialization:} Let $t=0$, each CU $i$ randomly chooses the
$\mathbf{a}^0(i)\in\mathcal{W}$ and
$\mathbf{p}^0_i\in\mathcal{F}_{i,\mathbf{a}^0(i)}$.\\
2) {\bf Update CU Memory:} For each $i\in\mathcal{N}$, talk to all
AP in the system and obtain $\mathbf{I}^t_i$. Shift
$\mathbf{a}^t(i)$, $\mathbf{I}^t_i$, and
$R_i\left(\mathbf{p}^t_{i,\mathbf{a}^t(i)}, \mathbf{p}^t_{-i,
\mathbf{a}^t(i)}; \mathbf{a}^t(i)\right)$ into the end of the AM, IM
and RM, respectively.
If $t>M$, shift the first element of the AM, IM and RM out of the memory.\\
3) {\bf Update AP Memory:} For each $w\in\mathcal{W}$, update the
vectors:
$\mathbf{p}({\mathcal{N}^t_w})=\left\{\mathbf{p}^t_{i,w}\right\}_{i\in\mathcal{N}^t_w}$;
$\mathbf{I}({\mathcal{N}^t_w})=\left\{\mathbf{I}^t_{i,w}\right\}_{i\in\mathcal{N}^t_w}$;
$\widehat{T}({\mathcal{N}^t_w})=\left\{
\begin{array}{ll}
\widehat{T}({\mathcal{N}^t_w})+1&\textrm{if}~
\widehat{T}({\mathcal{N}^t_w})\ne \infty\\
1&\textrm{if}~
\widehat{T}({\mathcal{N}^t_w})= \infty\\
\end{array} \right.$.\\
4) {\bf Sample Memory:} Let $\widehat{M}=\min\{M,t\}$, each CU $i$
uniformly samples its association memory:
\begin{align}
&\textrm{Sample}~
index^t_i=multi\Big([\underbrace{\frac{1}{\widehat{M}},\cdots,\frac{1}{\widehat{M}}}_{\widehat{M}~
elements}]\Big);\nonumber\\
&\textrm{Let}~
\widehat{{a}}^t_i=AM_i(index^t_i),~~\widehat{\mathbf{I}}^t_i=IM_i(index^t_i),
~~\widehat{R}^t_i=RM_i(index^t_i).\label{eqSampleMemory}
\end{align}
5) {\bf Calculate Best AP Association:} Each CU $i$ finds
association according to $\widehat{{a}}^t_i$,
$\widehat{\mathbf{I}}^t_i$ and $\widehat{R}^t_i$, i.e., find the set
of APs $\mathcal{W}_i^*$ such that:
\begin{align}
\mathcal{W}^*_i=
\left\{w:\max_{\mathbf{p}_{i,w}\in\mathcal{F}_{i,w}}
R_i\left(\mathbf{p}_{i,w},\widehat{\mathbf{I}}^t_{i,w};w\right)>\widehat{R}^t_i\right\}
\bigcup\widehat{{a}}^t_i.
\end{align}
Then randomly pick $w^*\in\mathcal{W}^*_i$, and set
$\mathbf{a}^{t+1}(i)=w^*$.\\
6) {\bf Calculate Power Allocation:} Each CU $i$ switches to AP
$\mathbf{a}^{t+1}(i)$. Let $w=\mathbf{a}^{t+1}(i)$, then CU $i$
obtain the following quantities from this AP $w$:
$\widehat{T}({\mathcal{N}^{t+1}_w})$,
$\mathbf{I}_i({\mathcal{N}^{t+1}_w})$, and
$\mathbf{p}_i({\mathcal{N}^{t+1}_w})$. If
$\widehat{T}({\mathcal{N}^{t+1}_w})\ge 1$ (the set of CUs
$\mathcal{N}_w$ has been associated with AP $w$ at the same time
before), let
$\widehat{\alpha}=\alpha_{\widehat{T}({\mathcal{N}^{t+1}_w})}$, and:
\begin{align}
\mathbf{p}^{t+1}_i=(1-\widehat{\alpha})\mathbf{p}_i({\mathcal{N}^{t+1}_w})+
\widehat{\alpha}\bfPhi_i\left(\mathbf{I}_i({\mathcal{N}^{t+1}_w})\right).\label{eqJJASPAPowerUpdate}
\end{align}
If $\widehat{T}({\mathcal{N}^{t+1}_w})= \infty$ (the set of CUs
$\mathcal{N}_w$ has not been associated with AP $w$ at the same time
before), randomly pick
$\mathbf{p}^{t+1}_i\in\mathcal{F}_{i,\mathbf{a}^{t+1}(i)}$.\\
7) {\bf Continue:} Let $t=t+1$, go to step 2).

We see that although algorithmically the J-JASPA is similar to the
previously introduced algorithms in the sense that the AP
associations are decided probabilistically, and the power profiles
are computed based on historical profiles and newly computed
components $\bfPhi(.)$, there are several significant differences
between the J-JASPA and the previously introduced
algorithms.\\
1) In J-JASPA, each CU calculates its { best AP association}
according to a {\it sampled historical} network state, while in the
JASPA and
Si-JASPA, it calculates this quantity according to the {\it current} network state.\\
2) In J-JASPA, CUs' AP association is the same as their {\it best AP
association}, while in Si-JASPA and JASPA, their AP
association is {\it sampled from the memory}. \\
3) This algorithm requires APs to have memory. Each AP needs to
record the local power allocation and interference profiles for {\it
all} the different sets of CUs that have been associated with it in
the previous iterations, while in the previously introduced
algorithms, the APs do not need to
have memory.\\
4) The J-JASPA requires larger memory for the CUs for constructing AM, IM and RM.\\
5) The J-JASPA requires extra communications between the CUs and the
APs ( mainly in step 6).

We will see in the next subsection that it is exactly these changes
in the algorithm and the extra requirements in terms of memory and
communication that enables the J-JASPA to have provable convergence
guarantees without the need to reach the intermediate equilibria.
This is a significant improvement compared with the original JASPA,
which does need the CUs to reach the intermediate equilibria, and
the Se/Si-JASPA, for which we are not able to provide complete
convergence proofs. We have also observed in simulation that J-JASPA
converges faster than Se/Si-JASPA.

\subsection{The convergence of the J-JASPA algorithm}
In this subsection, we show that the J-JASPA algorithm converges to
a JEP.

Define the set $\mathcal{A}$ as follows: $
\mathbf{a}\in\mathcal{A}\Longleftrightarrow\mathbf{a}~\textrm{
infinitely often in~}\{\mathbf{a}^t\}_{t=1}^{\infty}$. We first
provide a proposition characterizing the power profiles of the CUs
in the network every time a profile $\mathbf{a}\in\mathcal{A}$
appears.
\newtheorem{P7}{Proposition}
\begin{P1}\label{propIO2}
{\it Choose $\mathbf{a}\in\mathcal{A}$. Let
$\{t(n)\}_{n=1}^{\infty}$ be the subsequence of
$\{t\}_{t=1}^{\infty}$ such that $\mathbf{a}$ is played, i.e.,
$\left\{t(n):\mathbf{a}^{t(n)}=\mathbf{a}\right\}$. Then we have for
all $w\in\mathcal{W}$, $\lim_{n\to\infty}\mathbf{p}_w^{t(n)}=
\mathbf{p}_w^*(\mathbf{a})$, where $\mathbf{p}_w^*(\mathbf{a})$ is a
NE power allocation profile for AP $w$ under $\mathbf{a}$, i.e.,
$\mathbf{p}_w^*(\mathbf{a})\in\mathcal{E}_w(\mathbf{a})$.
Furthermore, we have that $
\lim_{n\to\infty}P_w\left(\mathbf{p}_w^{t(n)};
\mathbf{a}^{t(n)}\right)=\bar{P}_w(\mathbf{a}) $ and $
\lim_{n\to\infty}P\left(\mathbf{p}^{t(n)},\mathbf{a}^{t(n)}\right)=\bar{P}(\mathbf{a}).
$ }
\end{P1}
\begin{proof}
For a $w\in\mathcal{W}$, let $\mathcal{N}_w=\{i:\mathbf{a}(i)=w\}$.
Define another subsequence $\{\bar{t}(n)\}_{n=1}^{\infty}$ in which
the subset of CUs $\mathcal{N}_w$ is associated with AP $w$.
Clearly, $\{t(n)\}_{n=1}^{\infty}$ is a subsequence of
$\{\bar{t}(n)\}_{n=1}^{\infty}$. From the J-JASPA algorithm, we see
that at each $\bar{t}(n)$, \eqref{eqJJASPAPowerUpdate} implements
the single AP A-IWF (cf. Algorithm 1 in \cite{hong10m_japsa_1}) with
the {\it fixed} set of CUs $\mathcal{N}_w$. Thus, from Proposition 2
in \cite{hong10m_japsa_1} we have  that the subsequence
$\left\{\mathbf{p}^{\bar{t}(n)}_w\right\}_{n=1}^{\infty}$ converges
to $\mathbf{p}^*_w(\mathbf{a})\in\mathcal{E}_w(\mathbf{a})$, which
is a NE power allocation profile under fixed system association
profile $\mathbf{a}$. Consequently, the infinite sub-subsequence
$\left\{\mathbf{p}^{t(n)}_w\right\}_{n=0}^{\infty}$ also converges
to the same $\mathbf{p}^*_w(\mathbf{a})$. From Corollary
\ref{corollaryPotentialMaximization}, we have that
$\lim_{n\to\infty}P_w\left(\mathbf{p}_w^{t(n)};
\mathbf{a}^{t(n)}\right)=\bar{P}_w(\mathbf{a})$ and
$\lim_{n\to\infty}P\left(\mathbf{p}^{t(n)},\mathbf{a}^{t(n)}\right)=\bar{P}(\mathbf{a})$.
\end{proof}

We need the following set of definitions to proceed. Let
$\widehat{\mathbf{a}}^t$ be the {\it sampled system profile} at time
$t$:
$\widehat{\mathbf{a}}^t(i)\triangleq\widehat{a}^t_i,~\forall~i$. For
a specific $\mathbf{a}$, define the subsequence
$\{t(n,\mathbf{a})\}$ be the time instances that ${\mathbf{a}}$
appears and is immediately sampled by all the CUs, i.e.,
$\left\{t(n,\mathbf{a}): \mathbf{a}^{t(n,\mathbf{a})}=\mathbf{a}
~\textrm{and}~
\widehat{\mathbf{a}}^{t(n,\mathbf{a})}={\mathbf{a}}\right\}$. Note
that if $\mathbf{a}^t=\mathbf{a}$, then according to step 5) of the
J-JASPA algorithm, with positive probability
$\widehat{\mathbf{a}}^t=\mathbf{a}$. Thus, if
$\mathbf{a}\in\mathcal{A}$, then $\{t(n,\mathbf{a})\}$ is a infinite
sequence. Define $R^*_i\left(
\widehat{\mathbf{I}}^t_{i,w};w\right)\triangleq\max_{\mathbf{p}_{i,w}\in\mathcal{F}_{i,w}}
R_i\left(\mathbf{p}_{i,w}, \widehat{\mathbf{I}}^t_{i,w};w\right)$ to
be the maximum rate CU $i$ can achieve in AP $w$ based on sampled
interference $\widehat{\mathbf{I}}^t_{i,w}$. Define a set:
\begin{align}
B_i\left(\widehat{\mathbf{I}}^t_i,\widehat{\mathbf{a}}^t(i)\right)\triangleq
\left\{w:R^*_i\left( \widehat{\mathbf{I}}^t_{i,w};w\right)>
\widehat{R}^t_i\right\}\bigcup\widehat{\mathbf{a}}^t(i)
\end{align}
where $\widehat{R}^t_i$ is defined in \eqref{eqSampleMemory}. We
call the set
$B_i\left(\widehat{\mathbf{I}}^t_i,\widehat{\mathbf{a}}^t(i)\right)$
CU $i$'s {\it best association set at time $t$}. From step 5) of the
J-JASPA algorithm, all $w\in
B_i\left(\widehat{\mathbf{I}}^t_i,\widehat{\mathbf{a}}^t(i)\right)$
has positive probability to be picked by CU $i$ in iteration $t+1$.
Let
$B_i\left({\mathbf{I}}^*_i(\mathbf{a}),{\mathbf{a}}(i)\right)\triangleq
\lim_{n\to\infty}B_i\left(\widehat{\mathbf{I}}^{t(n,\mathbf{a})}_i,\widehat{\mathbf{a}}^{t(n,\mathbf{a})}(i)\right)$\footnote{Note
that such limit exist because of Proposition \ref{propIO2}.}. We
provide a technical characterization of the best association set.
See Appendix \ref{app3} for proof.


\newtheorem{P8}{Proposition}
\begin{P1}\label{propInclusion}
{\it For a specific CU $i$ and a system association profile
$\mathbf{a}\in\mathcal{A}$, suppose there exists a
$w\ne\mathbf{a}(i)$ such that $w \in
B_i\left({\mathbf{I}}^*_i(\mathbf{a}),{\mathbf{a}}(i)\right)$, i.e.,
CU $i$ has the incentive to move to a different AP in the limit.
Then there exists a large enough constant $N_i^*(\mathbf{a})$ such
that for all $n> N_i^*(\mathbf{a})$, we have:
\begin{align}
B_i\left({\mathbf{I}}^*_i(\mathbf{a}),{\mathbf{a}}(i)\right)
\subseteq
B_i\left(\widehat{\mathbf{I}}^{t(n,\mathbf{a})}_i,\widehat{\mathbf{a}}^{t(n,\mathbf{a})}(i)\right).\label{eqBestReplySetEqual}
\end{align}}
\end{P1}

In words, this propositions says that suppose a specific association
profile $\mathbf{a}$ happens infinitely often (hence is sampled
infinitely often), and suppose in the limit, when $\mathbf{a}$ is
sampled, a CU $i$ prefers a $w\ne\mathbf{a}(i)$, then after
iteration $t(N^*_i(\mathbf{a}),\mathbf{a})$, it must prefer $w$ in
every time instance $t(n,\mathbf{a})$, where $n\ge
N^*_i(\mathbf{a})$. Now we are ready to provide the main result for
the J-JASPA algorithm.

\newtheorem{T7}{Theorem}
\begin{T1}\label{theoremConvergenceJJASPA}
The J-JASPA algorithm converges to a JEP with probability 1.
\end{T1}
\begin{proof}
Consider the sequence $\{\left(\mathbf{a}^t,
\mathbf{p}^t\right)\}_{t=1}^{\infty}$. Choose
$\widetilde{\mathbf{a}}$ to be any system association profile that
satisfies the following: $
\widetilde{\mathbf{a}}\in\arg\max_{\mathbf{a}\in\mathcal{A}}\{\bar{P}(\mathbf{a})\}.
$ We first show that
$\left(\widetilde{\mathbf{a}},\mathbf{p}^*(\widetilde{\mathbf{a}})\right)$
is a JEP.

Suppose
$\left(\widetilde{\mathbf{a}},\mathbf{p}^*(\widetilde{\mathbf{a}})\right)$
is not a JEP, then there exists a CU $\check{i}$, and a
$\check{w}\ne\widetilde{\mathbf{a}}({\check{i}})$ such that
$\check{w}\in
B_{\check{i}}\left({\mathbf{I}}^*_{\check{i}}(\widetilde{\mathbf{a}}),{\widetilde{\mathbf{a}}}(i)\right)$.
This implies that there exists an $\underline{\epsilon}>0$ such
that:
\begin{align}
R^*_{\check{i}}\left(
{\mathbf{I}}^*_{{\check{i}},\check{w}}(\widetilde{\mathbf{a}});\check{w}\right)-R_{\check{i}}\left(
{\mathbf{I}}^*_{\check{i},\widetilde{\mathbf{a}}({\check{i}})}(\widetilde{\mathbf{a}});
\widetilde{\mathbf{a}}({\check{i}})\right)\ge \underline{\epsilon}.
\end{align}

%

Define a new association profile $\check{\mathbf{a}}$ as:
\begin{align}
\check{\mathbf{a}}=\left\{ \begin{array}{ll}
\widetilde{\mathbf{a}}(j) &\textrm{for j}~\ne~{\check{i}}\\
\check{w} & \textrm{for j}~={\check{i}.} \\
\end{array}\right.
\end{align}

Following the steps we already went through in Theorem
\ref{theoremExistence} from \eqref{eqAlphaContradiction} to
\eqref{eqContradiction}, we can show that:
\begin{align}
\bar{P}(\widetilde{\mathbf{a}})<\bar{P}(\check{\mathbf{a}}).
\end{align}
It is clear that if $\check{\mathbf{a}}\in\mathcal{A}$, then the
above is a contradiction to the assumption that $
\widetilde{\mathbf{a}}\in\arg\max_{\mathbf{a}\in\mathcal{A}}\{\bar{P}(\mathbf{a})\}
$. In the following, we show that
$\check{\mathbf{a}}\in\mathcal{A}$, thus completing the proof.

From Proposition \ref{propInclusion}, there exists a
$N_{\check{i}}^*(\widetilde{\mathbf{a}})$ large enough that for all
${n}>N_{\check{i}}^*(\widetilde{\mathbf{a}})$, $\check{w}\in
B_{\check{i}}\left(\widehat{\mathbf{I}}^{t(n,\widetilde{\mathbf{a}})}_{\check{i}},\widehat{\mathbf{a}}^{t(n,\widetilde{\mathbf{a}})}({\check{i}})\right)$.
Take any $n>N_{\check{i}}^*(\widetilde{\mathbf{a}})$. We know that
from the definition, in iteration $t(n,\widetilde{\mathbf{a}})$,
$\widehat{\mathbf{a}}^{t(n,\widetilde{\mathbf{a}})}(i)=\widetilde{\mathbf{a}},~\forall~i\in\mathcal{N}$.
From Step 5) in the J-JASPA algorithm, we see that with positive
probability, in iteration $t(n,\widetilde{\mathbf{a}})+1$, CU $j\ne
{\check{i}}$ chooses to stay in $\widetilde{\mathbf{a}}(j)$, and CU
${\check{i}}$ chooses to switch to $\check{w}$. This implies that
the association profile $\check{\mathbf{a}}$ happens with positive
probability in every time instance $t(n,\widetilde{\mathbf{a}})+1$.
Because $\{t(n,\widetilde{\mathbf{a}})\}$ is a infinite sequence,
$\check{\mathbf{a}}$ happens infinitely often, i.e.,
$\check{\mathbf{a}}\in\mathcal{A}$.


In summary, we conclude that $\widetilde{\mathbf{a}}$ must be a NE
association profile, and thus,
$\left(\widetilde{\mathbf{a}},\mathbf{p}^*(\widetilde{\mathbf{a}})\right)$
is a JEP.

Finally, following the proofs of Theorem
\ref{theoremConvergenceJASPA}, we can show similarly that the
sequence $\left\{(\mathbf{a}^t,
\mathbf{p}(\mathbf{a}^t))\right\}_{t=1}^{\infty}$ that produced by
J-JASPA converges to a JEP with probability 1.
\end{proof}

\section{Simulation Results}\label{secSimluation}
In this section, we present various simulation results to validate
the proposed algorithms. We first show the results regarding to the
convergence properties, and then present the results regarding to
the system throughput performance. Due to the space limit, for each
experiment we show the results obtained by running either
Si/Se-JASPA and J-JASPA, or the results obtained by the original
JASPA.

We have the following general settings for the simulation. We place
multiple CUs and APs randomly in a $10m\times10m$ area; we let
$d_{i,w}$ denote the distance between CU $i$ and AP $w$, then the
channel gains between CU $i$ and AP $w$,
$\{|h_{i,w}(k)|^2\}_{k\in\mathcal{K}_w}$, are independently drawn
from an exponential distribution with mean $\frac{1}{d^2_{i,w}}$
(i.e., $|h_{i,w}(k)|$ is assumed to have Rayleigh distribution).
We let the available channels to be evenly pre-assigned to different
APs. When we say a ``snapshot" of the network, we refer to the
network with fixed (but randomly generated as above) AP, CU
locations and channel gains. We set the length of the individual
memory as $M=10$. For ease of presentation and comparison, when we
use the JASPA algorithm with connection cost, we let all the CUs'
connection cost $\{c_i\}_{i\in\mathcal{N}}$ be identical.

\subsection{Convergence}
\begin{figure*}[ht] \vspace*{-.5cm}
    \begin{minipage}[t]{0.52\linewidth}
    \centering
    {\includegraphics[width=
1\linewidth]{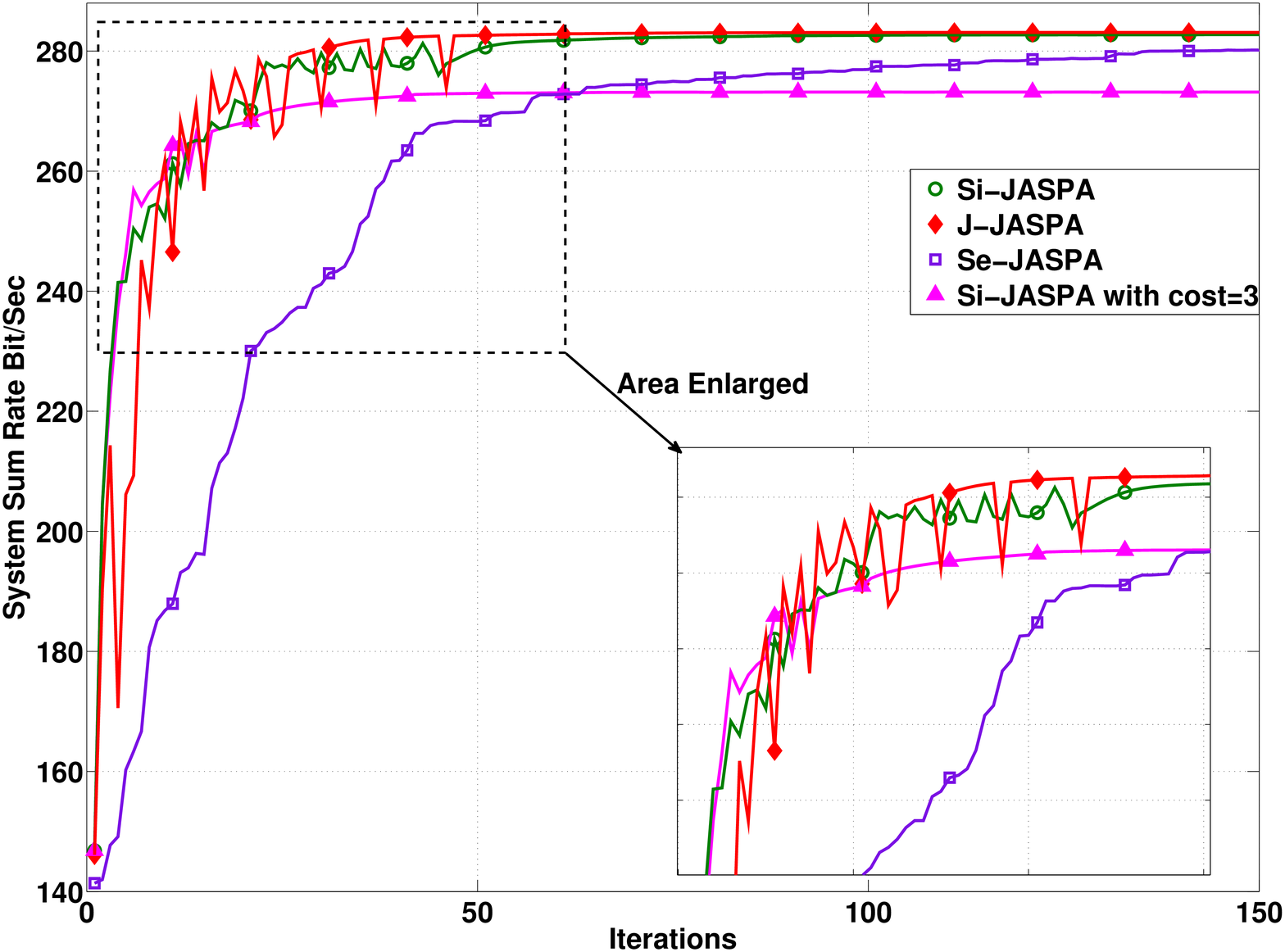} \vspace*{-0.5cm} \vspace*{-1cm}}
\end{minipage}
    \begin{minipage}[t]{0.52\linewidth}
    \centering
    {\includegraphics[width=
1\linewidth]{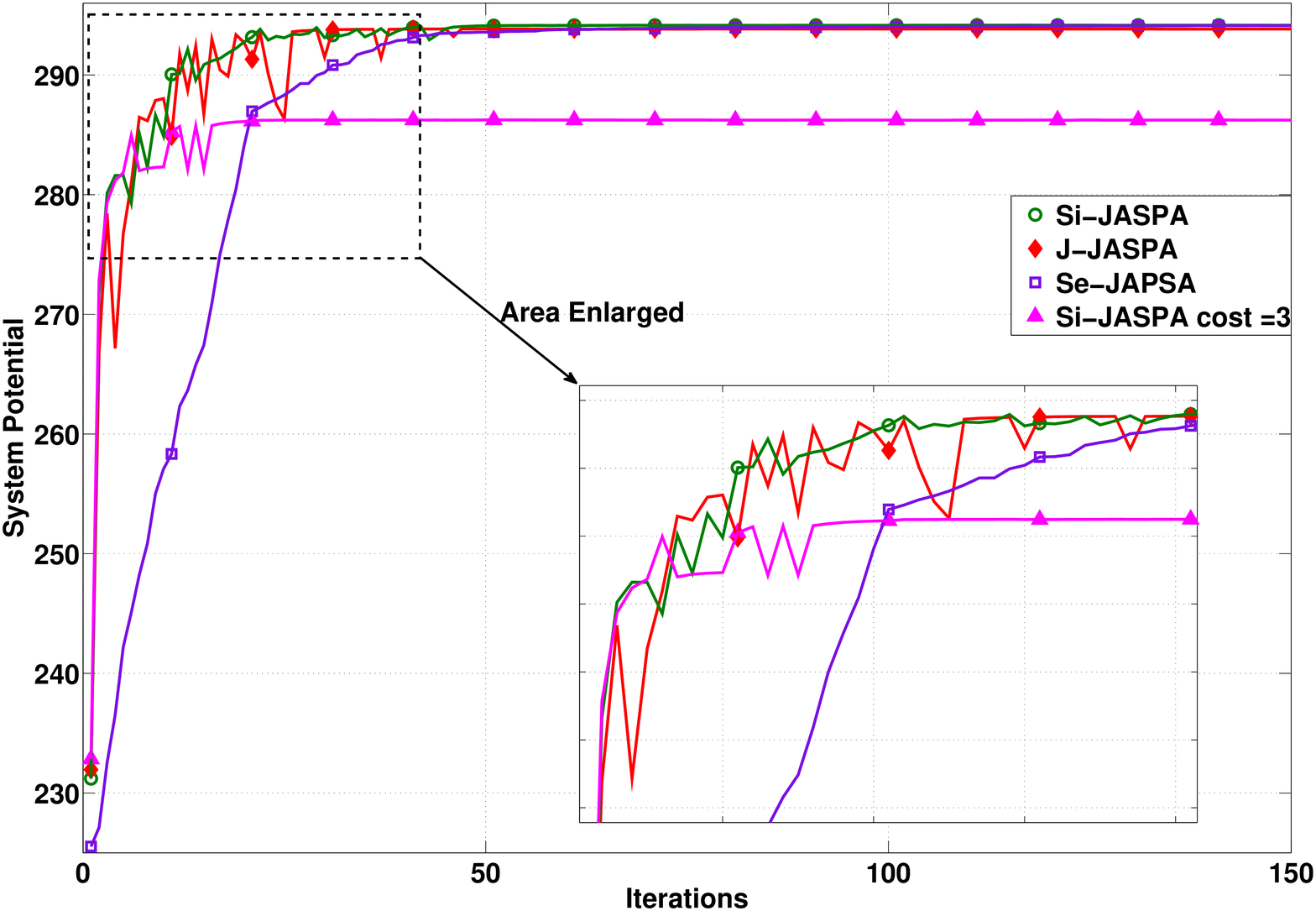} \vspace*{-0.5cm}
\vspace*{-1cm}}
\end{minipage}
\caption{Comparison of convergence speed by different algorithms.
Left: evolution of system sum rate. Right: evolution of the value of
system potential function.}\label{figCompareSpeed}\vspace*{-0.5cm}
    \end{figure*}

We only show the results for Si/Se-JASPA and J-JASPA in this
subsection. We first consider a network with $20$ CUs, $64$
channels, and $4$ APs. Fig. \ref{figCompareSpeed} shows the
evolution of the system throughput as well as the values of the
system potential function generated by a typical run of the
Se-JASPA, Si-JASPA, J-JASPA and Si-JASPA with connection cost
$c_i=3$ bit/sec $\forall~i\in\mathcal{N}$. We observe that the
Si-JASPA with connection cost converges faster than Si-JASPA and
Se-JASPA, while Se-JASPA converges very slowly. After convergence,
the system throughput achieved by Si-JASPA with connection cost is
smaller than that of the other three algorithms. Notice that in the
right part of Fig. \ref{figCompareSpeed}, the system potential
generated by the Se-JASPA is non-decreasing along iterations. This
property has been identified in Theorem
\ref{theoremConvergenceSeJASPA}.
%

%

\begin{figure*}[htb] \vspace*{-.3cm}
        \begin{minipage}[t]{0.33\linewidth}
    \centering
    {\includegraphics[width=
1\linewidth]{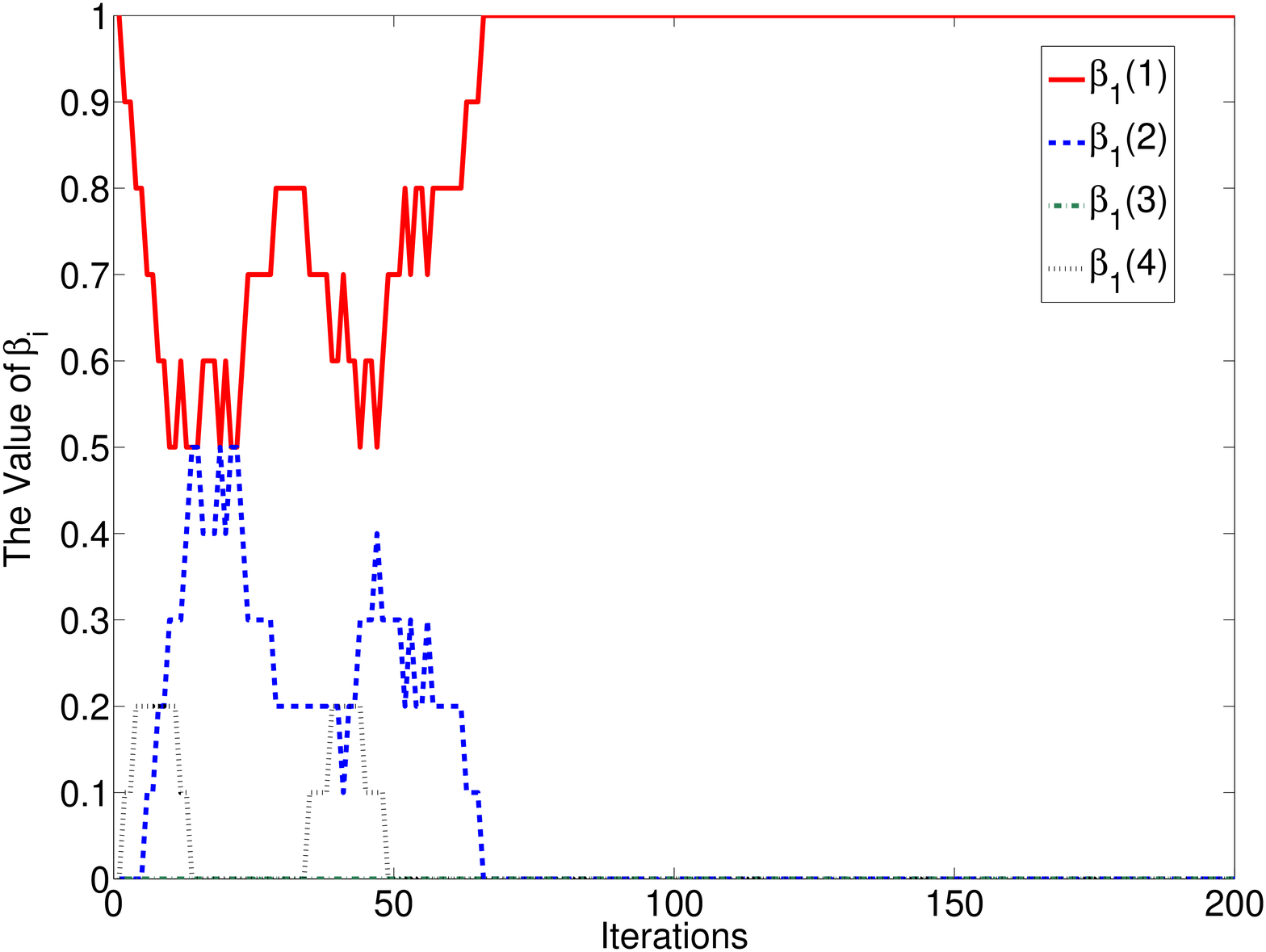}
\vspace*{-.3cm} \vspace*{-.8cm}}
\end{minipage}
\begin{minipage}[t]{0.33\linewidth}
    \centering
    {\includegraphics[width=
1\linewidth]{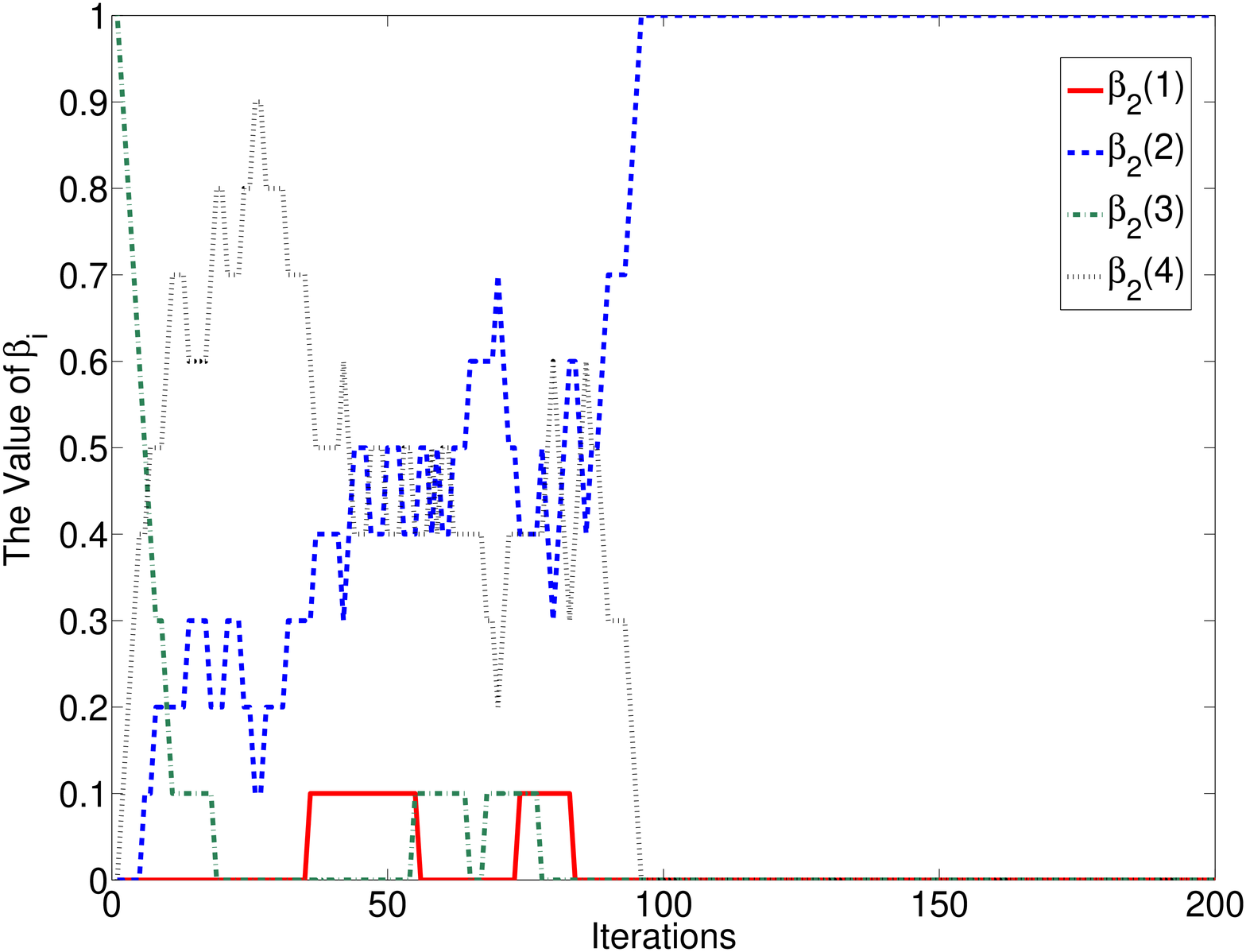}
\vspace*{-.3cm}\vspace*{-.8cm}}
\end{minipage}
    \begin{minipage}[t]{0.33\linewidth}
    \centering
    {\includegraphics[width=
1\linewidth]{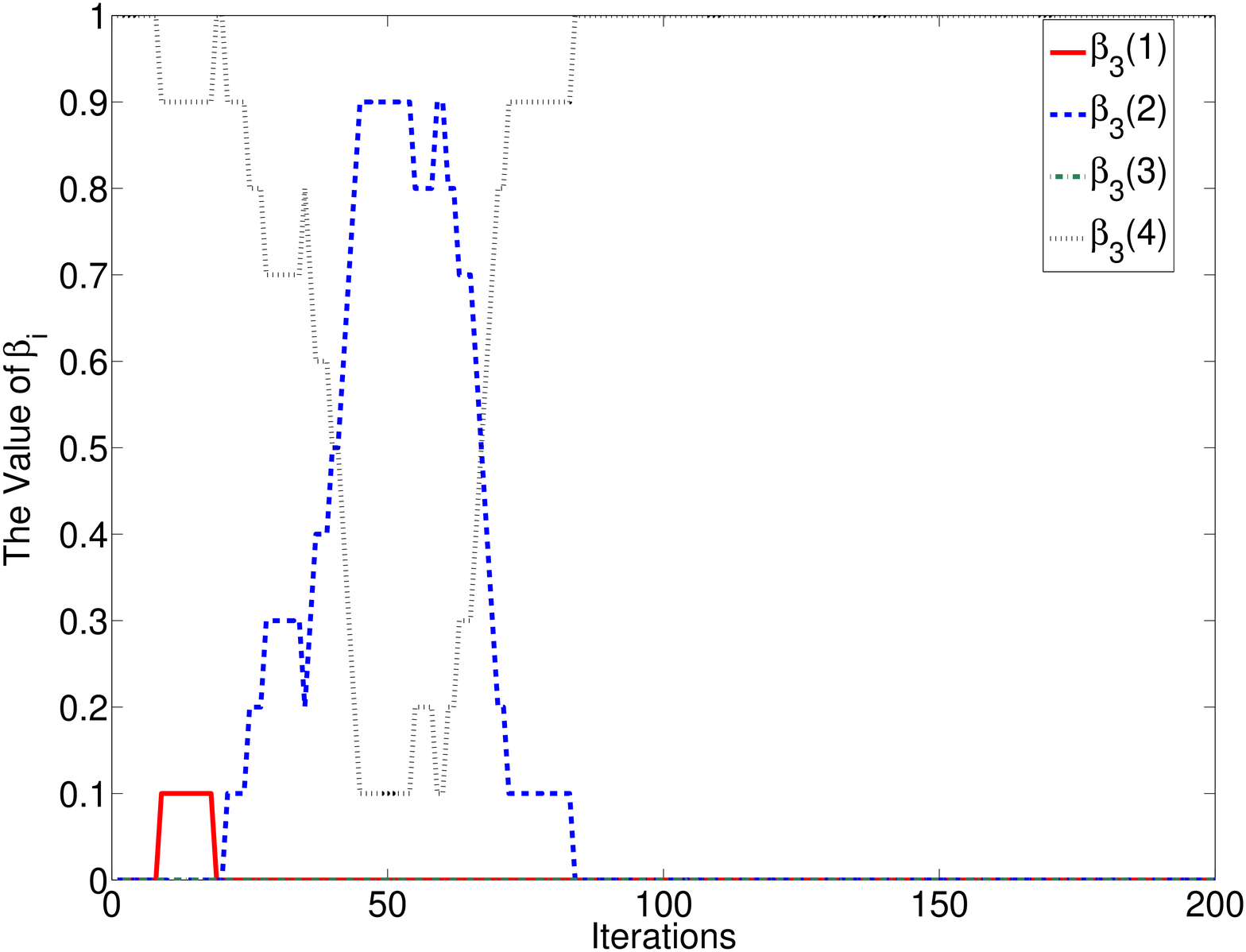}
\vspace*{-0.8cm}}
\end{minipage}
\caption{Convergence of the probability vector $\beta^t_1$,
$\beta^t_2$ and $\beta^t_3$ of CUs' 1, 2,
3.}\label{figConvergenceBeta}\vspace*{-0.5cm}
    \end{figure*}

\begin{figure*}[ht] \vspace*{-.5cm}

    \begin{minipage}[t]{0.5\linewidth}
    \centering
    {\includegraphics[width=
1\linewidth]{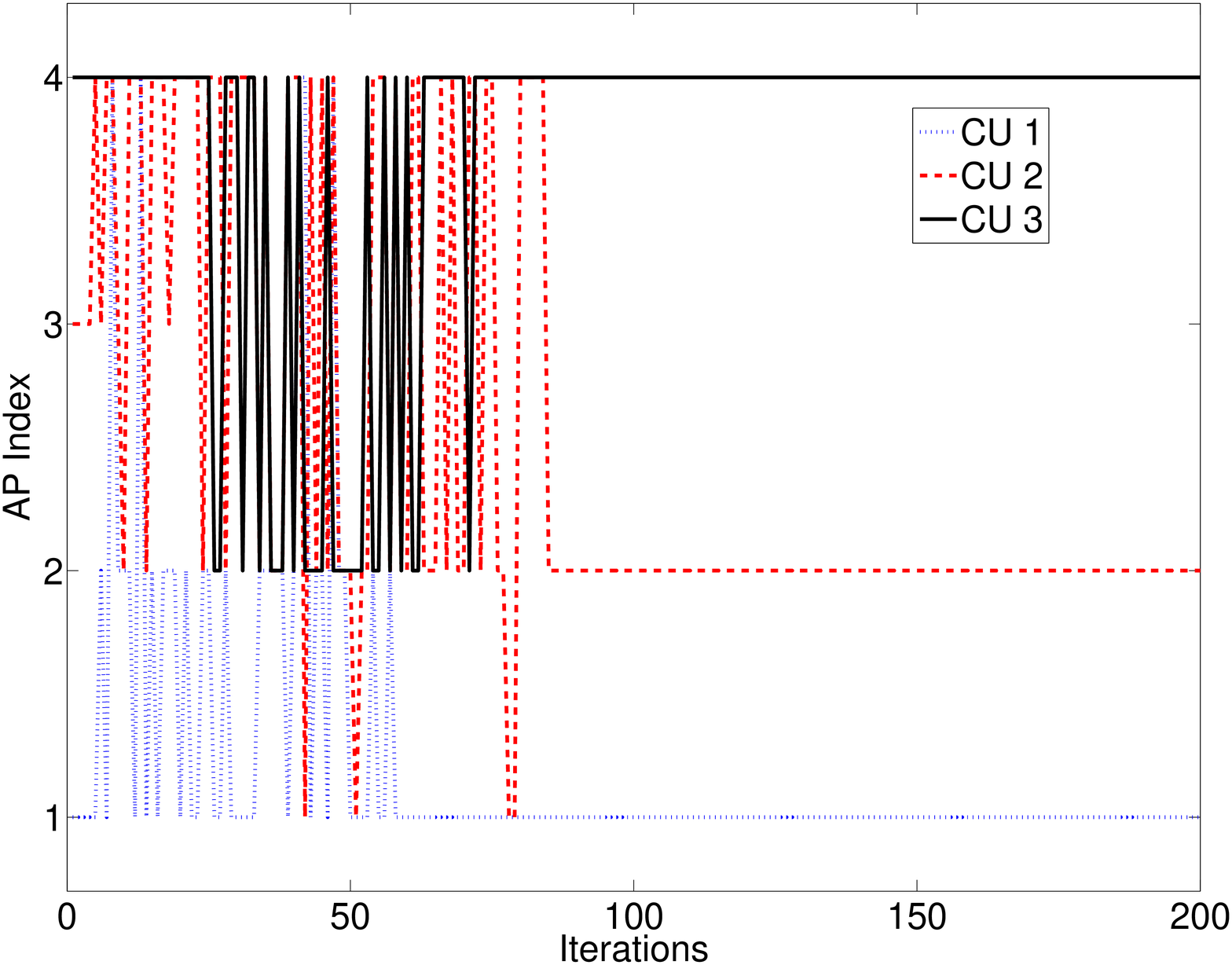}
\vspace*{-0.5cm}\caption{Convergence of Selected CUs' AP
selection.}\label{figConvergenceSelection} \vspace*{-1cm}}
\end{minipage}
    \begin{minipage}[t]{0.5\linewidth}
    \centering
    {\includegraphics[width=
1\linewidth]{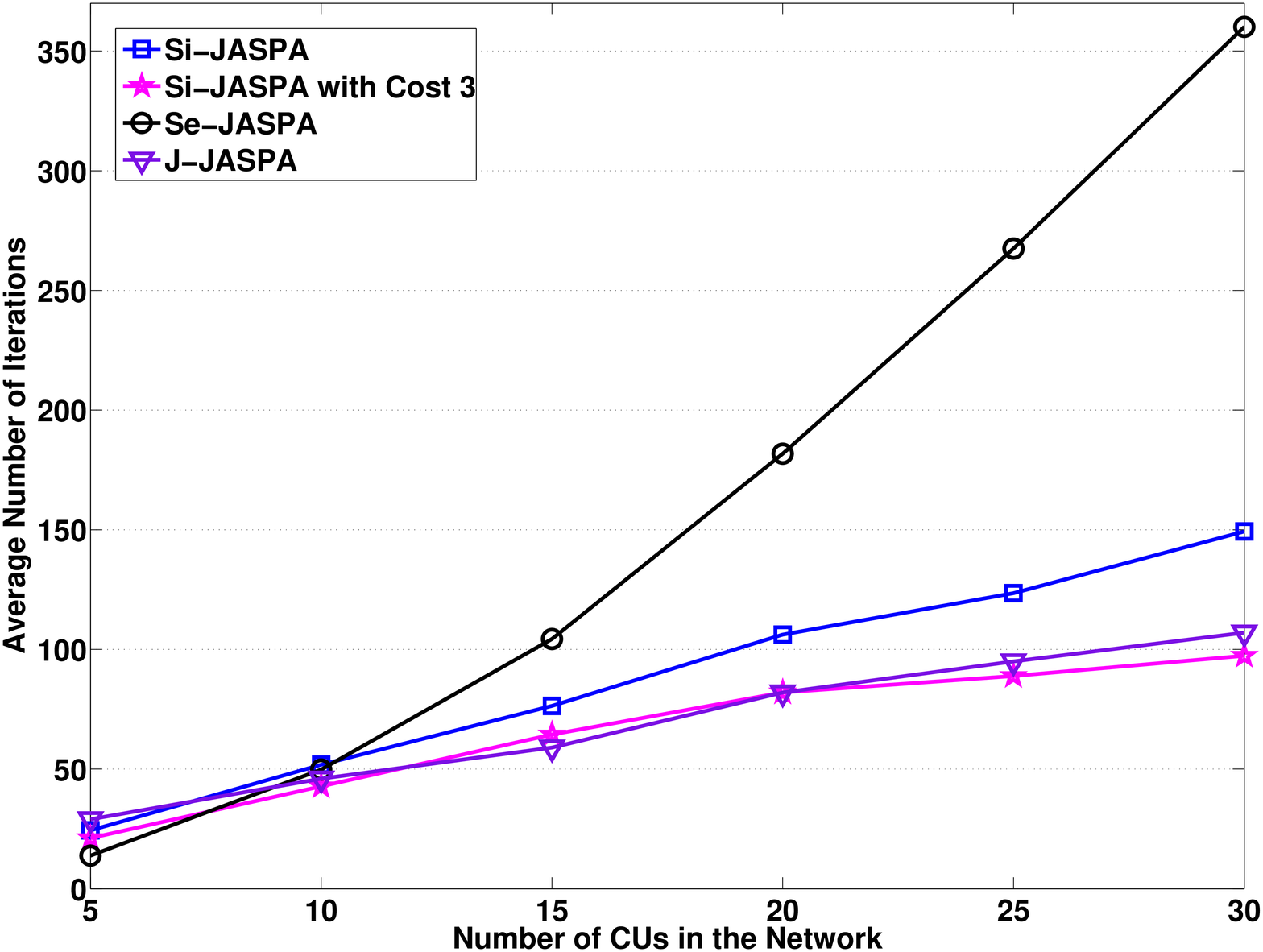}
\vspace*{-0.5cm}\caption{Comparison of averaged convergence
speeds.}\label{figConvergenceSpeed} \vspace*{-0.5cm}}
\end{minipage}

    \end{figure*}

Fig. \ref{figConvergenceSelection} shows the evolution of the AP
selections made by the CUs in the network during a typical run of
the Si-JASPA algorithm. We only show 3 out of 20 CUs (we refer the
selected CUs as CU 1, 2, 3 for easy reference) in order not to make
the figure overly crowded. Fig.\ref{figConvergenceBeta} shows the
corresponding evolution of the probability vectors
$\{\bfbeta^t_i\}_{t=1}^{200}$ for the three of the CUs selected in
Fig. \ref{figConvergenceSelection}. It is clear that upon
convergence, all the probability vector converges to a 0-1 vector.
We then evaluate how the number of CUs in the network affects the
speed of convergence of different algorithms. In order to do so, we
compare the average iterations to achieve convergence in the network
with 4 APs, 64 channels and different number of CUs, for the
following three algorithms 1) Si-JASPA, 2) Se-JASPA, 3) J-JASPA, 4)
Si-JASPA with connection cost $c_i=3$ bit/sec for all CUs. From
Fig.\ref{figConvergenceSpeed}, we see  that when the number of CUs
in the system becomes large, the sequential version of the JASPA
takes significantly longer time to converge than the other three
simultaneous versions of the JASPA algorithm. Moreover, the J-JASPA
shows faster convergence than the Si/Se-JASPA. We can also see that
the connection costs adopted by individual CUs indeed have positive
effects on the convergence speed of the system.

Note that each point in this figure represents the average of 100
independent runs of each algorithm on randomly generated network
snapshots.

\subsection{System Throughput Performance}\label{subThroughput}
We then evaluate the network throughput performance achievable by
the JEP computed by the JASPA.

We first investigate a small networks with $8$ CUs, $64$ channels
and $1,~2,~3,~4$ APs, and compare the performance of JASPA related
algorithms to the maximum network throughput that can be achieved
for the same network. The maximum network throughput for a snapshot
of the network is calculated by the following two steps: 1) for a
specific AP-CU association profile, say $\mathbf{a}$, calculate the
maximum network throughput (denoted by $T(\mathbf{a})$) by summing
up the maximum capacity\footnote{For a single AP with fixed number
of users and channel gains, the maximum capacity is the well-known
multiple access channel sum capacity.} of individual APs in the
network; 2) enumerate {\it all possible} AP-CU association profiles,
and find $T^*=\max_{\mathbf{a}}T(\mathbf{a})$. It is clear now that
the reason we choose to focus on such relatively small networks in
this experiment is that for a large network, the time it takes for
the above exhaustive search procedure to find the maximum network
throughput becomes prohibitive.

\begin{figure*}[ht] \vspace*{-.5cm}
    \begin{minipage}[t]{0.5\linewidth}
    \centering
    {\includegraphics[width=
1\linewidth]{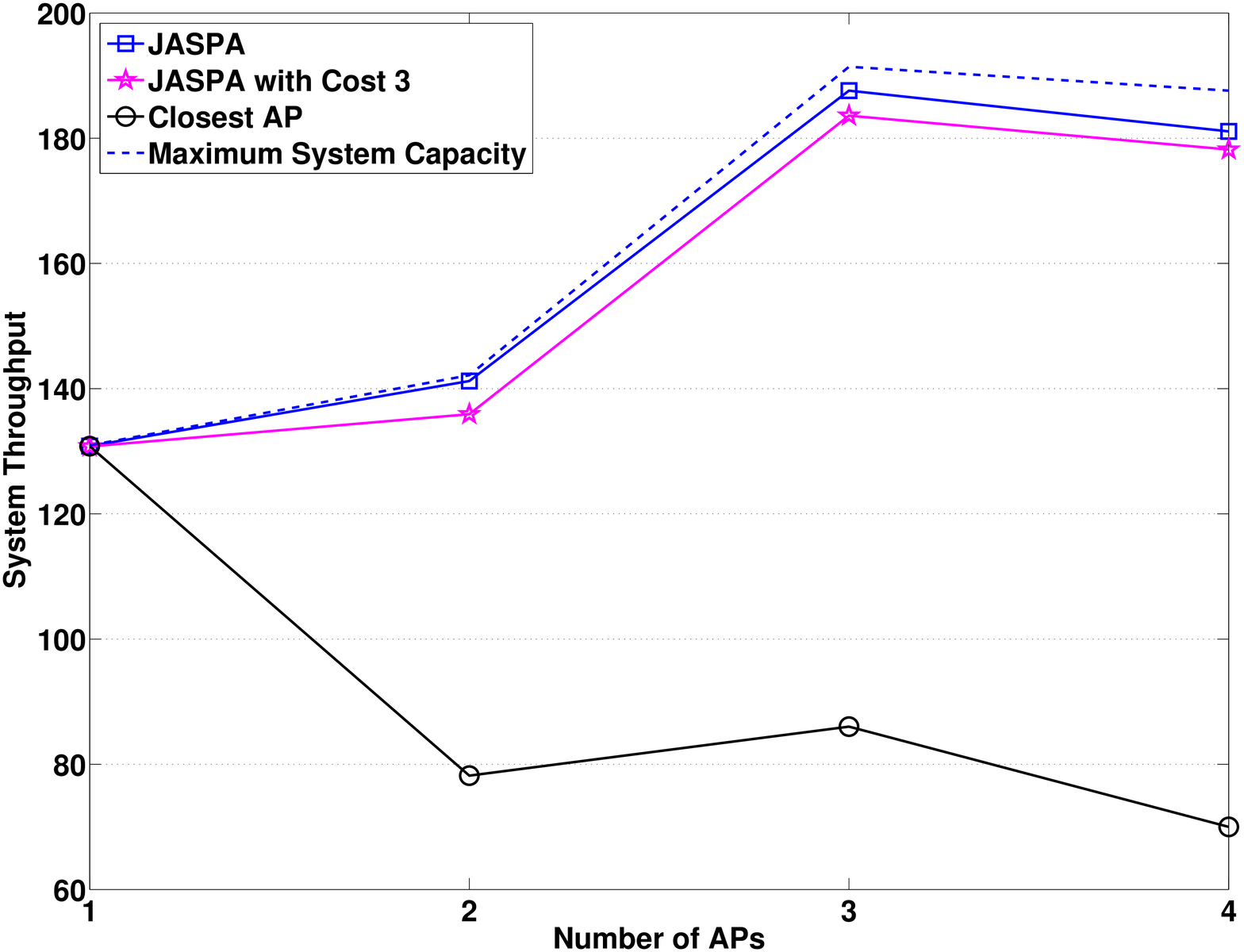}
\vspace*{-0.5cm}\caption{Comparison of the system throughput by
different algorithms with the throughput upper bound in a 8 CU
network.}\label{figCompareThrouputSmallScale} \vspace*{-1cm}}
\end{minipage}
    \begin{minipage}[t]{0.5\linewidth}
    \centering
    {\includegraphics[width=
1\linewidth]{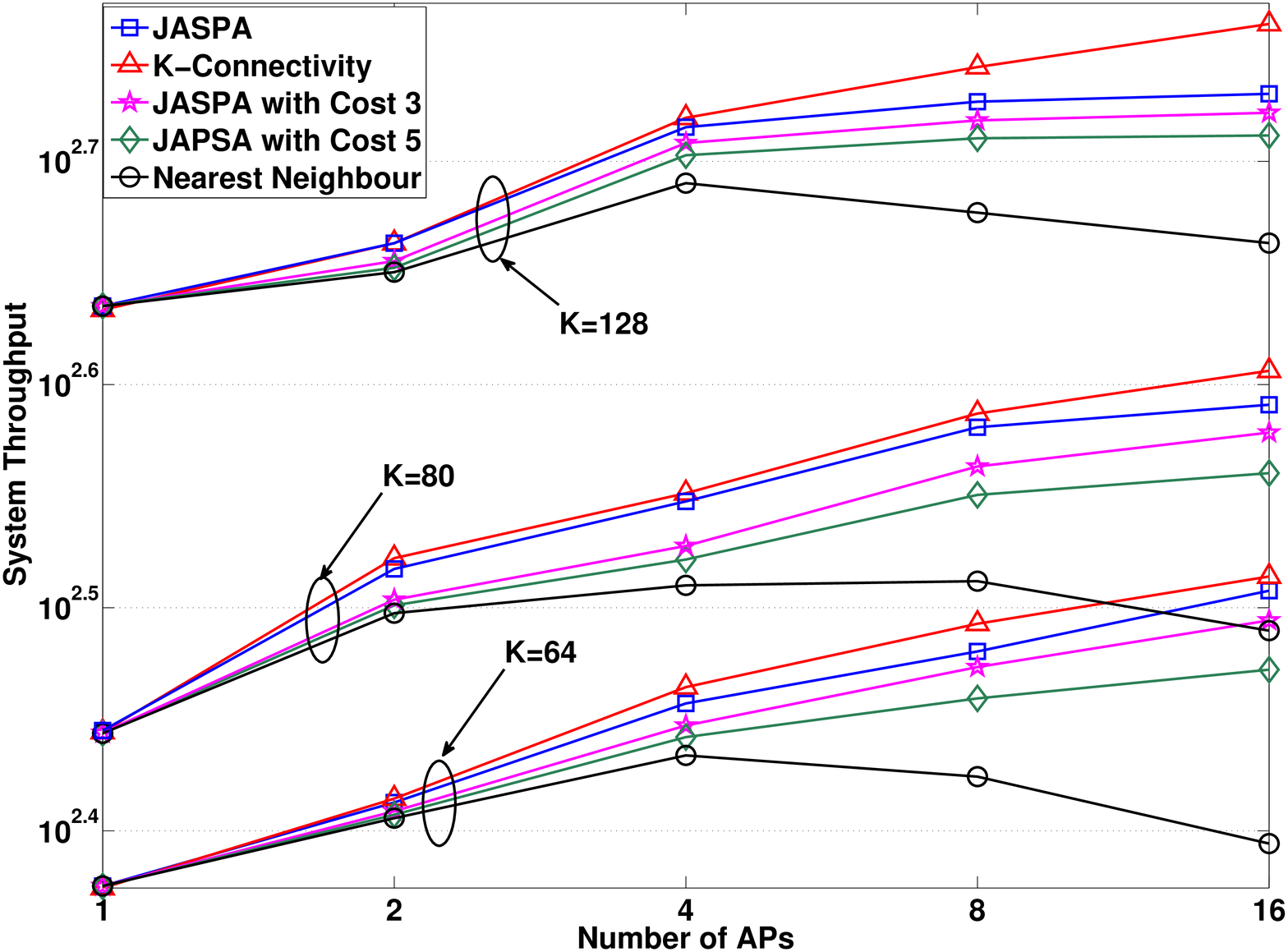}
\vspace*{-0.5cm}\caption{Comparison of the system throughput
\newline in a 30 CU network.}\label{figCompareThrouput} \vspace*{-0.5cm}}
\end{minipage}
    \end{figure*}

The result is shown in Fig.\ref{figCompareThrouputSmallScale}, where
each point on the figure is obtained by running the algorithms on
100 independent snapshots of the network. We see that the JASPA
algorithm performs very well with little throughput loss, while the
closest AP algorithm, which separates the tasks of spectrum decision
and spectrum sharing, performs poorly.


We then start to look at the performance of larger networks with 30
CUs, up to 16 APs and up to 128 channels. Fig.
\ref{figCompareThrouput} shows the comparison of the performance of
JASPA, JASPA with individual cost $c_i=3$ bit/sec and $c_i=5$
bit/sec, and the closest AP algorithm mentioned in section
\ref{subNearestAP}. We adopt the actual distance as the measure of
``closeness" in the closest AP algorithm. Each point in this figure
is the average of 100 independent runs of the algorithms.

Due to the prohibitive computation time required, we are unable to
obtain the maximum system throughput for these relatively large
networks. We instead compute the equilibrium system throughput that
can be achieved in a game {\it if all CUs are able to connect to
multiple APs at the same time}. We refer to this as the {\it
multiple-connectivity} network. It is clear that in such network,
there is no need for the CUs to perform the AP selection, and the
CUs in this network enjoy the flexibility of being able to connect
to multiple APs at the same time. However, we observe that the
performance of JASPA is close to that of the
``multiple-connectivity" network.
%

From Fig. \ref{figCompareThrouput} we see that when the number of
APs increases, the throughput of the JASPA algorithm becomes much
better than the closest AP algorithm, a phenomenon that is partly
due to the fact that for the closest AP algorithm, the separation of
the AP selection and power allocation process results in the
insufficient use of the spectrum: when the number of AP increases,
it becomes increasingly more probable that several APs are idle
because no CUs are close to them. Fig. \ref{figCompareThrouput},
along with Fig. \ref{figConvergenceSpeed}, also serve to confirm our
early speculation that algorithms with connection cost can indeed
improve the convergence speed while reducing the system throughput.

We also observe from Fig.\ref{figCompareThrouput} that generally the
system throughput increases as the number of APs increases, which
suggests that the scheme that partitions the available spectrum and
assigns them to the APs with different geographical locations is
indeed more favorable than the scheme which uses a single AP to
manage all the spectrum. This phenomenon can be explained partly by
reasoning as follows: when using a single AP, it is likely that many
of the CUs are located far away from the AP, and thus none of their
channels have good quality; on the other hand, when using multiple
APs, although each CU can only use part of the available channels,
it is more probable that there is one or more APs that are located
in its vicinity, thus is able to provide good channel quality. We
have to mention here that, although not shown in
Fig.\ref{figCompareThrouput}, placing too many APs in the network
may also result in reduced network throughput as the chance of idle
APs increases as the number of AP increases (an extreme case is that
the number of APs is larger than the number of CUs).

\section{Conclusion}\label{secConclusion}
In this paper, we addressed the joint AP association and power
allocation problem in a CRN. We formulate the problem into a
non-cooperative game with hybrid strategy space. We characterized
the NE of this game, and provided distributed algorithms to reach
such equilibrium. Empirical evidence gathered from simulation
experiments suggests that the equilibrium has very promising quality
in term of the system throughput.

There can be many future extensions to this work. First of all, the
non-cooperative game with hybrid strategy space analyzed in this
paper can be applied to many other problems as well, for example,
the CRN with interference channel and segmented spectrum mentioned
at the end of section \ref{subMotivation}. Secondly, for the problem
considered in this work, it is beneficial to characterize
quantitatively the efficiency of the JEP, and to provide solutions
for efficiency improvement. Thirdly, it will be interesting to
analyze the effect of time-varying channel gains and the arrival and
departure of the CUs on the performance of the algorithm, and to
propose suitable heuristic dealing with these situations.

\appendices
\section{Proof of Proposition \ref{propIO}}\label{app1}
\begin{proof}
Choose $\mathbf{c}\in\mathcal{C}$. and suppose that at time $t$,
$\mathbf{c}^t=\mathbf{c}$. If $\mathbf{c}$ is an equilibrium
association profile, then with probability at least
$(\frac{1}{M})^N$ (all CUs chooses
$\mathbf{a}^{t}(i)=\mathbf{c}^{t}(i)$), we have that $\mathbf{a}^{t}$ is
an equilibrium association profile. Because
$\mathbf{c}^t=\mathbf{c}$ happens infinitely often, we must have
$\mathbf{a}^{t}=\mathbf{c}$ happens infinitely often, i.e., there
exists an equilibrium association in the set $\mathcal{A}$.

Suppose $\mathbf{c}$ is not an equilibrium association. Then
consider the following steps of operation.

{\bf Step 1):} With probability at least $(\frac{1}{M})^N$,
$\mathbf{a}^t=\mathbf{c}$, and $\mathbf{c}\in\mathcal{A}$. Because
$\mathbf{c}$ is not a NE, then without loss of generality, assume
that CU $i$ is better off by switching to $\widehat{w}_i$:
$\mathbf{c}^{t+1}(i)=\widehat{w}_i\ne\mathbf{c}^{t}(i)$. Then we
must have that
$\bar{P}(\mathbf{a}^t)<\bar{P}\left((\widehat{w}_i,\mathbf{a}_{-i}^t)\right)$,
a fact from \eqref{eqContradiction}. Then with probability at least
$(\frac{1}{M})^N$ (all player $j$ except player $i$ choose
$\mathbf{a}^{t+1}(j)=\mathbf{c}^{t}(j)=\mathbf{a}^t(j)$) that
$\mathbf{a}^{t+1}=(\widehat{w}_i,\mathbf{a}_{-i}^t)$, and we have
$\bar{P}(\mathbf{a}^{t+1})>\bar{P}(\mathbf{a}^{t})$. Put index $i$
in the the set $\mathcal{U}$ : $\mathcal{U}=\{i\}$. We note in this
stage, we have:$ \mathbf{a}^{t+1}(i)=\mathbf{c}^{t+1}(i)$.
Similarly, at $t+2$, if we are able to find a CU $j\ne i$ with
$\mathbf{c}^{t+2}(j)=\widehat{w}_{j}\ne\mathbf{a}^{t+1}(j)$ (i.e. CU
$j$ is better off if switching to AP $\widehat{w}_{j}$), we let
$\mathcal{U}=\{i,j\}$. Then again with positive probability, we have
$\mathbf{a}^{t+2}=(\widehat{w}_{j},~\mathbf{a}_{-j}^{t+1})$.
Consequently, $\bar{P}(\mathbf{a}^{t+2})>\bar{P}(\mathbf{a}^{t+1})$.
We note in this stage, the following is true: $
\mathbf{a}^{t+2}(j)=\mathbf{c}^{t+2}(j)$. Continue this process,
until we reach a time $t+n<=t+N$ such that only CUs in the set
$\mathcal{U}$ are willing to switch. Let $\mathcal{E}$ be the
complement set of $\mathcal{U}$.

{\bf Step 2):} We must have that for
$j\in\mathcal{E}$, $\mathbf{a}^{t+n-1}(j)=\mathbf{c}^{t+n}(j)$.
On the other hand, for all $i\in\mathcal{U}$, from the argument in Step 1), we see that
there must exist a $0<k_i<n$ such that
$\mathbf{c}^{t+k_i}(i)=\mathbf{a}^{t+n-1}(i)$. Pick
$q\in\mathcal{U}$ such that $q=\arg\min_{i\in\mathcal{U},
\mathbf{c}^{t+n}(i)\ne\mathbf{a}^{t+n-1}(i)}k_i$. Consequently, we can
shift $\mathbf{c}^{t}$ out of the memory and still be able to
construct $\mathbf{a}^{t+n}=(\widehat{w}_q,\mathbf{a}_{-q}^{t+n-1})$
with positive probability, because all the elements in
$\mathbf{a}_{-q}^{t+n-1}$ must have been appeared once in
$\{\mathbf{c}^t\}^{t+n}_{t=t+1}$.  Move $q$ out of $\mathcal{U}$ and
into $\mathcal{E}$, and continue Step 2) until only CUs in the set
$\mathcal{E}$ are willing to switch. Switch the role of
$\mathcal{U}$ and $\mathcal{E}$, and continue Step 2).

By continuously performing the above operations,
$\{\bar{P}(\mathbf{a}^{t})\}_{t+1}^{\infty}$ is a strictly
increasing sequence, and there must exist a {\it finite} time
instance $T<\infty$ such that it is not possible to find
$\mathbf{a}^{t+T+1}$ that differs with $\mathbf{a}^{t+T}$ with a
single element and has the property
$\bar{P}(\mathbf{a}^{t+T+1})>\bar{P}(\mathbf{a}^{t+T})$.  Consequently, $\mathbf{a}^{t+T}$ is an
equilibrium profile. We let $\mathbf{a}^*=\mathbf{a}^{t+T}$.The
finiteness of $T$ comes from the finiteness of the number of values
of $\bar{P}(\mathbf{a})$ (due to the finiteness of the choice of
$\mathbf{a}$). Such finiteness combined with the strict positivity of the
probability of performing each operation in Step 1) and Step 2) implies that the probability of
reaching $\mathbf{a}^*$
from $\mathbf{c}$ is non-zero.

We conclude from the above analysis that with {\it positive}
probability, a NE profile $\mathbf{a}^*$ will appear after
$\mathbf{a}^t$ in {\it finite} steps. Because
$\mathbf{a}^t=\mathbf{c}$ happens infinitely often, we must also
have that $\mathbf{a}^*$ happens infinitely often, i.e.,
$\mathbf{a}^*\in\mathcal{A}$.

Finally, it is straightforward to see that the fact that
$\mathbf{a}^*$ is an equilibrium association profile suggests that
the tuple $(\mathbf{a}^*,\mathbf{p}^*(\mathbf{a}^*))$ is a JEP.
\end{proof}

\section{Proof of Theorem
\ref{theoremConvergenceSeJASPA}}\label{app2}
\begin{proof}
Suppose that at time $t+1$, it is CU $i$'s turn to move. Let
$w=\mathbf{a}^t(i)$ be the CU $i$'s associated AP at time $t$. We have the following two situations.\\
1) {\bf~At time $t+1$, CU $i$ is best off switching to
$\widehat{w}\ne\mathbf{a}^t(i)$}. In this situation, CU $i$'s
communication rate at time $t$ under association profile
$\mathbf{a}^t$ is as follows:{\small
\begin{align}
&{R}_i({\mathbf{p}}^t_{i,{w}},\mathbf{p}^t_{-i,{w}};w)=\sum_{k\in\mathcal{K}_{{w}}}
\log\left(1+\frac{|h_{i,{w}}(k)|^2
{p}^t_{i,{w}}(k)}{n_{{w}}(k)+\sum_{j\ne i:
,\mathbf{a}^t(j)={w}}|h_{j,{w}}(k)|^2 p^t_{j,{w}}(k)}\right)\nonumber\\
&\stackrel{(a)}=P_{{w}}(\mathbf{p}^t_{{w}};{{\mathbf{a}}^t})-P_{{w}}(\mathbf{p}_{-i,{w}}^{t};{\mathbf{a}^{t+1}})
\stackrel{(b)}=P_{{w}}(\mathbf{p}^t_{{w}};{{\mathbf{a}}^t})-P_{{w}}(\mathbf{p}_{{w}}^{t+1};{\mathbf{a}^{t+1}})
\end{align}}
where $(a)$ and $(b)$ are true because of the fact that at time
$t+1$, CU $i$ no longer associates with AP $w$, and all other CUs
keep their power and association profile the same as in time $t$.

CU $i$'s communication rate after it finishes switching to
$\widehat{w}$ is:{\small
\begin{align}
&{R}_i({\mathbf{p}}^{t+1}_{i,\widehat{w}},\mathbf{p}^t_{\widehat{w}};{\widehat{w}})=\sum_{k\in\mathcal{K}_{\widehat{w}}}
\log\left(1+\frac{|h_{i,\widehat{w}}(k)|^2
{p}^{t+1}_{i,\widehat{w}}(k)}{n_{\widehat{w}}(k)+\sum_{j:
,\mathbf{a}^t(j)=\widehat{w}}|h_{j,\widehat{w}}(k)|^2 p^t_{j,\widehat{w}}(k)}\right)\nonumber\\
&=P_{\widehat{w}}({\mathbf{p}}^{t+1}_i,\mathbf{p}^t_{\widehat{w}};{{\mathbf{a}}^{t+1}})-P_{\widehat{w}}(\mathbf{p}_{\widehat{w}}^t;{\mathbf{a}^t})=P_{\widehat{w}}(\mathbf{p}^{t+1}_{\widehat{w}};{{\mathbf{a}}^{t+1}})-P_{\widehat{w}}(\mathbf{p}_{\widehat{w}}^t;{\mathbf{a}^t}).\label{eqEstimatedRate2}
\end{align}}
Because $
{R}_i({\mathbf{p}}^{t+1}_{i,\widehat{w}},\mathbf{p}^t_{\widehat{w}};{\widehat{w}})>
{R}_i({\mathbf{p}}^t_{i,{w}},\mathbf{p}^t_{-i,{w}};{{w}})$, we
have:{\small
\begin{align}
P_{\widehat{w}}(\mathbf{p}^{t+1}_{\widehat{w}};{{\mathbf{a}}^{t+1}})+
P_{{w}}(\mathbf{p}_{{w}}^{t+1};{\mathbf{a}^{t+1}})>
P_{{w}}(\mathbf{p}^t_{{w}};{{\mathbf{a}}^t})+P_{\widehat{w}}(\mathbf{p}_{\widehat{w}}^t;{\mathbf{a}^t}).\nonumber
\end{align}}
Arguing similarly as in \eqref{eqSumPotentialInequality}, we have
that SEP must satisfy: $
P(\mathbf{p}^{t+1};{\mathbf{a}^{t+1}})>P(\mathbf{p}^{t};{\mathbf{a}^{t}}).
$

2){\bf~At Time $t+1$, CU $i$ stays in ${w}$}. Notice that in this
case, we have $\mathbf{a}^{t+1}=\mathbf{a}^t$. From Proposition 3 of
\cite{hong10m_japsa_1}, we have the following inequality:{\small $
P_w(\mathbf{p}^{t+1}_{i,w},\mathbf{p}^{t+1}_{-i,w};{\mathbf{a}^{t+1}})=
P_w(\mathbf{p}^{t+1}_{i,w},\mathbf{p}^{t}_{-i,w};{\mathbf{a}^t}) \ge
P_w(\mathbf{p}^{t}_{i,w},\mathbf{p}^{t}_{-i,w};{\mathbf{a}^t})\nonumber
$}, thus, $ P(\mathbf{p}^{t+1};{\mathbf{a}^{t+1}})\ge
P(\mathbf{p}^{t};{\mathbf{a}^{t}}). $ We conclude that in both
cases, the system potential is non-decreasing. Because
$P(\mathbf{p}^{t};{\mathbf{a}^{t}})$ is upper bounded,
$\left\{P(\mathbf{p}^{t};{\mathbf{a}^{t}})\right\}_{t=1}^{\infty}$
is a converging sequence. 
%
\end{proof}

\section{Proof of Proposition \ref{propInclusion}}\label{app3}
\begin{proof}
From Proposition \ref{propIO2}, if $\mathbf{a}\in\mathcal{A}$, we
have that
$\lim_{n\to\infty}\mathbf{p}^{t(n,\mathbf{a})}_w=\mathbf{p}_w^*(\mathbf{a}),\forall~i\in\mathcal{N}$,
which implies that
$\lim_{n\to\infty}\mathbf{I}_{i,w}^{t(n,\mathbf{a})}=\mathbf{I}_{i,w}^*(\mathbf{a}),
\forall~i\in\mathcal{N}$. This result combined with the continuity
of the function $R^*_i\left(
\widehat{\mathbf{I}}^{t(n,\mathbf{a})}_{i,w};w\right)$ with respect
to $\widehat{\mathbf{I}}^{t(n,\mathbf{a})}_{i,w}$, and the
continuity of the function
$\widehat{R}^{t(n,\mathbf{a})}_i=R_i\left(\mathbf{p}^{t(n,\mathbf{a})}_w,
\mathbf{a}^{t(n,\mathbf{a})}(i)\right)$ with respect to
$\mathbf{p}^{t(n,\mathbf{a})}_w$, further implies that, for any
$\delta>0$, there must be a $N(\delta)$ such that for all $n
>N(\delta)$, the followings are true:{\small
\begin{align}
\max_{w\in\mathcal{W}}\left|R^*_i\left(
\widehat{\mathbf{I}}^{t(n,\mathbf{a})}_{i,w};w\right)-R^*_i\left(
{\mathbf{I}}^*_{i,w}(\mathbf{a}); w\right)\right|&<\delta,\nonumber\\
\left|\widehat{R}^{t(n,\mathbf{a})}_i-{R}_i(\mathbf{p}^*(\mathbf{a});\mathbf{a}(i))\right|&<\delta.
\end{align}}
For any $w\ne\mathbf{a}(i)$ such that $w\in
B_i\left({\mathbf{I}}^*_i(\mathbf{a}),{\mathbf{a}}\right)$, there
must exits a $\epsilon_w>0$ such that:{\small
\begin{align}
R^*_i\left( {\mathbf{I}}^*_{i,w}(\mathbf{a});w\right)-R_i\left(
{\mathbf{I}}^*_{i,\mathbf{a}(i)}(\mathbf{a});\mathbf{a}(i)\right)\ge
{\epsilon}_w.
\end{align}}
Take $\epsilon>0$ such that $\epsilon=\min_{w\in
B_i\left({\mathbf{I}}^*_i(\mathbf{a}),{\mathbf{a}}(i)\right)}\epsilon_w$,
and choose a $\widehat{\delta}$ small enough such that
$0<2\widehat{\delta}<{\epsilon}$, and let
$N_i^*(\mathbf{a})\triangleq N(\widehat{\delta})$. We have that for
all $n>N_i^*(\mathbf{a})$, the following is true:{\small
\begin{align}
&R^*_i\left( {\mathbf{I}}^*_{i,w}(\mathbf{a});w\right)-R_i\left(
{\mathbf{I}}^*_{i,\mathbf{a}(i)}(\mathbf{a});\mathbf{a}(i)\right)\nonumber\\
&= R_i\left( {\mathbf{I}}^*_{i,w}(\mathbf{a});w\right) + R^*_i\left(
\widehat{\mathbf{I}}^{t(n,\mathbf{a})}_{i,w};w\right)-R^*_i\left(
\widehat{\mathbf{I}}^{t(n,\mathbf{a})}_{i,w};w\right) +
\widehat{R}^{t(n,\mathbf{a})}_i-\widehat{R}^{t(n,\mathbf{a})}_i
 - R_i\left(
{\mathbf{I}}^*_{i,\mathbf{a}(i)}(\mathbf{a});\mathbf{a}(i)\right)\nonumber\\
&\le R^*_i\left(
\widehat{\mathbf{I}}^{t(n,\mathbf{a})}_{i,w};w\right)-\widehat{R}^{t(n,\mathbf{a})}_i+
\left|R^*_i\left(
{\mathbf{I}}^*_{i,w}(\mathbf{a});w\right)-R^*_i\left(
\widehat{\mathbf{I}}^{t(n,\mathbf{a})}_{i,w};w\right)
\right|+\left|\widehat{R}^{t(n,\mathbf{a})}_i- R_i\left(
{\mathbf{I}}^*_{i,\mathbf{a}(i)}(\mathbf{a});\mathbf{a}(i)\right)\right|\nonumber\\
&\le R^*_i\left(
\widehat{\mathbf{I}}^{t(n,\mathbf{a})}_{i,w};w\right)-\widehat{R}^{t(n,\mathbf{a})}_i+\widehat{\delta}+\widehat{\delta}.
\end{align}}
Consequently, we have that for all $n>N_i^*(\mathbf{a})$, $
R^*_i\left(\widehat{\mathbf{I}}^{t(n,\mathbf{a})}_{i,w};w\right)-\widehat{R}^{t(n,\mathbf{a})}_i
\ge {\epsilon}-2\widehat{\delta}>0, $ which implies that $w$ must be
in the set
$B_i\left(\widehat{\mathbf{I}}^{t(n,\mathbf{a})}_i,\widehat{\mathbf{a}}^{t(n,\mathbf{a})}(i)\right)$.
The claim is proved.
\end{proof}

\bibliographystyle{IEEEbib}
\bibliography{ref}

\begin{thebibliography}{10}

\bibitem{hong11_infocom}
M.~Hong, A.~Garcia, and J.~Barrera,
\newblock ``Joint distributed {AP} selection and power allocation in cognitive
  radio networks\,''
\newblock in {\em the Proceedings of the IEEE INFOCOM}, 2011,
\newblock accepted.

\bibitem{akyildiz08}
I.~F. Akyildiz, W.~Y. Lee, M.~C. Vuran, and S.~Mohanty,
\newblock ``A survey on spectrum management in cognitive radio networks,''
\newblock {\em IEEE Communications Magazine}, pp. 40--48, April 2008.

\bibitem{lai08}
L.~Lai and H.~E. Gamal,
\newblock ``The water-filling game in fading multiple-access channels,''
\newblock {\em IEEE Transactions on Information Theory}, vol. 54, no. 5, 2008.

\bibitem{Meshkati06}
F.~Meshkati, M.~Chiang, H.~V. Poor, and S.~C. Schwartz,
\newblock ``A game-theoretic approach to energy-efficient power control in
  multicarrier {CDMA} systems,''
\newblock {\em IEEE Journal on Selected Areas in Communications}, vol. 24, pp.
  1115--1129, 2006.

\bibitem{islam08}
M.~H. Islam, Y.-C. Liang, and A.~T. Hoang,
\newblock ``Joint power control and beamforming for cognitive radio networks,''
\newblock {\em IEEE Transactions on Wireless Communications}, vol. 7, no. 7,
  pp. 2415--2419, 2008.

\bibitem{stevenson09}
C.~R. Stevenson, G.~Chouinard, Z.~Lei, W.~Hu, S.~J. Shellhammer, and
  W.~Caldwell,
\newblock ``{IEEE} 802.22: the first cognitive radio wireless regional area
  network standard,''
\newblock {\em Comm. Mag.}, vol. 47, no. 1, pp. 130--138, 2009.

\bibitem{song05a}
G.~Song and Y.~Li,
\newblock ``Cross-layer optimization for {OFDM} wireless networks--part {I}:
  Theoretical framework,''
\newblock {\em IEEE Transactions on Wireless Communications}, vol. 4, no. 2,
  pp. 614--624, 2005.

\bibitem{song05b}
G.~Song and Y.~Li,
\newblock ``Cross-layer optimization for {OFDM} wireless networks--part {II}:
  Algorithm development,''
\newblock {\em IEEE Transactions on Wireless Communications}, vol. 4, no. 2,
  pp. 625--634, 2005.

\bibitem{luo04}
Z-.Q. Luo, T.~N. Davidson, G.~B. Giannakis, and K.~M. Wong,
\newblock ``Transceiver optimization for block-based multiple access through
  {ISI} channels,''
\newblock {\em IEEE Transactions on Signal Processing}, vol. 52, no. 4, pp.
  1037--1052, 2004.

\bibitem{yu02b}
W.~Yu and J.~M. Cioffi,
\newblock ``{FDMA} capacity of gaussian multiple-access channel with isi,''
\newblock {\em IEEE Transactions on Communications}, vol. 50, no. 1, pp.
  102--111, 2002.

\bibitem{kim05b}
K.~Kim, Y.~Han, and S.-L Kim,
\newblock ``Joint subcarrier and power allocation in uplink {OFDMA} systems,''
\newblock {\em IEEE Communication Letters}, vol. 9, pp. 526--528, 2005.

\bibitem{liu10}
T.~Liu, C.~Yang, and L.-L. Yang,
\newblock ``A lower-complexity subcarrier-power allocation scheme for
  frequency-division multiple-access scheme,''
\newblock {\em IEEE Transactions on Wireless Communications}, vol. 11, no. 5,
  pp. 1571--1576, 2010.

\bibitem{Li07}
H.~Li and H.~Liu,
\newblock ``An analysis of uplink {OFDM} optimality,''
\newblock {\em IEEE Transactions on Wireless Communications}, vol. 6, no. 8,
  pp. 2972--2983, 2007.

\bibitem{he08}
G.~He, S.~Gault, M.~Debbah, and E.~Altman,
\newblock ``Distributed power allocation game for uplink ofdm systems,''
\newblock in {\em Proc. WiOPT}, 2008, pp. 515--521.

\bibitem{acharya09}
J.~Acharya and R.~D. Yates,
\newblock ``Dynamic spectrum allocation for uplink users with heterogeneous
  utilities,''
\newblock {\em IEEE Transactions on Wireless Communications}, vol. 8, no. 3,
  pp. 1405--1413, 2009.

\bibitem{yu04}
W.~Yu, W.~Rhee, S.~Boyd, and J.~M. Cioffi,
\newblock ``Iterative water-filling for gaussian vector multiple-access
  channels,''
\newblock {\em IEEE Transactions on Information Theory}, vol. 50, no. 1, pp.
  145--152, 2004.

\bibitem{monderer96}
D.~Monderer and L.~S. Shapley,
\newblock ``Potential games,''
\newblock {\em Games and Economics Behaviour}, vol. 14, pp. 124--143, 1996.

\bibitem{scutari08a}
G.~Scutari, D.~P. Palomar, and S.~Barbarossa,
\newblock ``Optimal linear precoding strategies for wideband noncooperative
  systems based on game theory -- part {I}: Nash equilibria,''
\newblock {\em IEEE Transactions on Signal Processing}, vol. 56, no. 3, 2008.

\bibitem{luo06b}
Z-.~Q. Luo and J-.S. Pang,
\newblock ``Analysis of iterative waterfilling algorithm for multiuser power
  contorl in digital subscriber lines,''
\newblock {\em EURASIP Journal on Applied Signal Processing}, vol. 2006, pp.
  1--10, 2006.

\bibitem{zhao07}
Q.~Zhao and B.~M. Sadler,
\newblock ``A survey of dynamic spectrum access,''
\newblock {\em IEEE Signal Processing Magazine}, , no. 5, pp. 79--89, 2007.

\bibitem{cover05}
T.~M. Cover and J.~A. Thomas,
\newblock {\em Elements of Information Theory, second edition},
\newblock Wiley, 2005.

\bibitem{osborne94}
M.~J. Osborne and A.~Rubinstein,
\newblock {\em A Course in Game Theory},
\newblock MIT Press, 1994.

\bibitem{deb08}
R.~Deb,
\newblock ``A characterization of differentiable potential games,''
\newblock {\em http://www.econ.yale.edu/~rd287/}.

\bibitem{scutari06}
G.~Scutari, S.~Barbarossa, and D.~P. Palomar,
\newblock ``Potential games: A framework for vector power control problems with
  coupled constraints,''
\newblock in {\em the Proceedings of ICASSP 06}, 2006.

\bibitem{cheng93}
R.~Cheng and S.~Verdu,
\newblock ``Gaussian multiaccess channels with isi: Capacity region and
  multiuser water-filling,''
\newblock {\em IEEE Transactions on Information Theory}, vol. 39, no. 3, pp.
  773--785, 1993.

\bibitem{yu02a}
W.~Yu, G.~Ginis, and J.~M. Cioffi,
\newblock ``Distributed multiuser power control for digital subscriber lines,''
\newblock {\em IEEE Journal on Selected Areas in Communications}, vol. 20, no.
  5, pp. 1105--1115, 2002.

\bibitem{scutari08b}
G.~Scutari, D.~P. Palomar, and S.~Barbarossa,
\newblock ``Optimal linear precoding strategies for wideband noncooperative
  systems based on game theory -- part {II}: Algorithms,''
\newblock {\em IEEE Transactions on Signal Processing}, vol. 56, no. 3, 2008.

\bibitem{shum07}
K.~W. Shum, K.~K. Leung, and C.~W. Sung,
\newblock ``Convergence of iterative waterfilling algorithm for gaussian
  interference channels,''
\newblock {\em IEEE Journal on Selected Area in Communications}, vol. 25, pp.
  1091--1100, 2007.

\bibitem{bertsekas97}
D.~P. Bertsekas and J.~N. Tsitsiklis,
\newblock {\em Parallel and Distributed Computation: Numerical Methods, 2nd
  ed},
\newblock Athena Scientific, Belmont, MA, 1997.

\bibitem{Ermoliev69}
Y.~M. Ermoliev,
\newblock ``On the method of generalized stochastic gradient and
  quasi-fej\'{e}r sequences,''
\newblock {\em Cybernetics}, 1969.

\bibitem{iusem94}
A.~N. Iusem, B.~F. Svaiter, and M.~Teboulle,
\newblock ``Entropy-like proximal methods in convex programming,''
\newblock {\em Mathematics of Operations Research}, 1994.

\bibitem{Burachik95}
R.~Burachik, L.~M.~G. Drummond, and A.~N. Iusem,
\newblock ``Full convergence of the steepest descent method with inexact line
  search,''
\newblock {\em Optimization}, pp. 137--146, 1995.

\bibitem{jindal05}
N.~Jindal, W.~Rhee, S.~Vishwanath, S.~A. Jafar, and A.~Goldsmith,
\newblock ``Sum power iterative water-filling for multi-antenna gaussian
  broadcast channels,''
\newblock {\em IEEE Transactions on information theory}, vol. 51, no. 4, 2005.

\bibitem{zhang08}
J.~Zhang, D.~Zhang, and M.~Chiang,
\newblock ``The impact of stochastic noisy feedback on distributed network
  utility maximization,''
\newblock {\em IEEE Transactions on Information Theory}, , no. 2, pp. 645--665,
  2008.

\bibitem{goldsmith05}
A.~Goldsmith,
\newblock {\em Wireless Communications},
\newblock Combridge University Press, New York, 2005.

\end{thebibliography}


\begin{thebibliography}{10}

\bibitem{hong11_infocom}
M.~Hong, A.~Garcia, and J.~Barrera,
\newblock ``Joint distributed {AP} selection and power allocation in cognitive
  radio networks\,''
\newblock in {\em the Proceedings of the IEEE INFOCOM}, 2011,
\newblock accepted.

\bibitem{hong10m_japsa_1}
M.~Hong, A.~Garcia, and S.~G. Wilson,
\newblock ``Distributed uplink resource allocation in cognitive radio
  networks--part {I}: Equilibria and algorithms for power allocation,''
\newblock manuscript in preparation.

\bibitem{stevenson09}
C.~R. Stevenson, G.~Chouinard, Z.~Lei, W.~Hu, S.~J. Shellhammer, and
  W.~Caldwell,
\newblock ``{IEEE} 802.22: the first cognitive radio wireless regional area
  network standard,''
\newblock {\em Comm. Mag.}, vol. 47, no. 1, pp. 130--138, 2009.

\bibitem{acharya09}
J.~Acharya and R.~D. Yates,
\newblock ``Dynamic spectrum allocation for uplink users with heterogeneous
  utilities,''
\newblock {\em IEEE Transactions on Wireless Communications}, vol. 8, no. 3,
  pp. 1405--1413, 2009.

\bibitem{apcan06}
T.~Alpcan and T.~Basar,
\newblock ``A hybrid noncooperative game model for wireless communications,''
\newblock {\em Annals of the International Society of Dynamic Games}, vol. 9,
  pp. 411--429, 2007.

\bibitem{saraydar01}
C.~U. Sarayda, N.~B. Mandayam, and D.~J. Goodman,
\newblock ``Pricing and power control in a multicell wireless data network,''
\newblock {\em IEEE Journal on selected areas in communications}, vol. 19, no.
  10, pp. 1883--1892, 2001.

\bibitem{hanly95}
S.~V. Hanly,
\newblock ``An algorithm for combined cell-site selection and power control to
  maximize cellular spread spectrum capacity,''
\newblock {\em IEEE Journal on selected areas in communications}, vol. 13, no.
  7, pp. 1332--1340, 1995.

\bibitem{yates95b}
R.~D. Yates and C.~Y. Huang,
\newblock ``Integrated power control and base station assignment,''
\newblock {\em IEEE Transactions on Vehicular Technology}, vol. 44, pp.
  1427--1432, 1995.

\bibitem{Meshkati06}
F.~Meshkati, M.~Chiang, H.~V. Poor, and S.~C. Schwartz,
\newblock ``A game-theoretic approach to energy-efficient power control in
  multicarrier {CDMA} systems,''
\newblock {\em IEEE Journal on Selected Areas in Communications}, vol. 24, pp.
  1115--1129, 2006.

\bibitem{shakkottai07}
S.~Shakkottai, E.~Altman, and A.~Kumar,
\newblock ``Multihoming of users to access points in {WLAN}s: A population game
  perspective,''
\newblock {\em IEEE Journal on Selected Areas In Communications}, , no. 6, pp.
  1207--1215, August 2007.

\bibitem{kauffmann07}
B.~Kauffmann, F.~Baccelli, and A.~Chaintreau,
\newblock ``Measurement-based self organization of interfering 802.11 wireless
  access network,''
\newblock in {\em the Proceedings of IEEE INFOCOM}, 2007, pp. 1451--1459.

\bibitem{bejerano07}
Y.~Bejerano, S.~J. Han, and L.~Li,
\newblock ``Fairness and load balancing in wireless lans using association
  control,''
\newblock {\em IEEE/ACM Transactions on Networking}, vol. 15, no. 3, pp.
  560--573, 2007.

\bibitem{kumar05}
A.~Kumar and V.~Kumar,
\newblock ``Optimal association of stations and aps in an ieee 802.11 wlan,''
\newblock in {\em Proceedings of NCC}, 2005.

\bibitem{kumar07}
A.~Kumar, E.~Altman, D.~Miorandi, and M.~Goyal,
\newblock ``New insights from a fixed-point analysis of single cell ieee 802.11
  wlans,''
\newblock {\em IEEE/ACM Trans. Netw.}, vol. 15, no. 3, pp. 588--601, 2007.

\bibitem{wang08}
F.~Wang, M.~Krunz, and S.~G. Cui,
\newblock ``Price-based spectrum management in cognitive radio networks,''
\newblock {\em IEEE Journal of Selected Topics in Signal Processing}, vol. 2,
  no. 1, 2008.

\bibitem{pang09}
J-.~S. Pang, G.~Scutari, D.~P. Palomar, and F.~Facchinei,
\newblock ``Design of cognitive radio systems under temperature-interference
  constraints: A variational inequality approach,''
\newblock {\em IEEE Transactions on Signal Processing},
\newblock Accepted for Publication.

\bibitem{wu09}
Y.~Wu and D.~H.~K. Tsang,
\newblock ``Distributed power allocation algorithm for spectrum sharing
  cognitive radio networks with qos guarantee,''
\newblock in {\em Proceedings of INFOCOM}, 2009.

\bibitem{yu02a}
W.~Yu, G.~Ginis, and J.~M. Cioffi,
\newblock ``Distributed multiuser power control for digital subscriber lines,''
\newblock {\em IEEE Journal on Selected Areas in Communications}, vol. 20, no.
  5, pp. 1105--1115, 2002.

\bibitem{yu04}
W.~Yu, W.~Rhee, S.~Boyd, and J.~M. Cioffi,
\newblock ``Iterative water-filling for gaussian vector multiple-access
  channels,''
\newblock {\em IEEE Transactions on Information Theory}, vol. 50, no. 1, pp.
  145--152, 2004.

\bibitem{luo06b}
Z-.~Q. Luo and J-.S. Pang,
\newblock ``Analysis of iterative waterfilling algorithm for multiuser power
  contorl in digital subscriber lines,''
\newblock {\em EURASIP Journal on Applied Signal Processing}, vol. 2006, pp.
  1--10, 2006.

\bibitem{scutari08a}
G.~Scutari, D.~P. Palomar, and S.~Barbarossa,
\newblock ``Optimal linear precoding strategies for wideband noncooperative
  systems based on game theory -- part {I}: Nash equilibria,''
\newblock {\em IEEE Transactions on Signal Processing}, vol. 56, no. 3, 2008.

\bibitem{scutari08b}
G.~Scutari, D.~P. Palomar, and S.~Barbarossa,
\newblock ``Optimal linear precoding strategies for wideband noncooperative
  systems based on game theory -- part {II}: Algorithms,''
\newblock {\em IEEE Transactions on Signal Processing}, vol. 56, no. 3, 2008.

\bibitem{cao10}
L.~Cao, L.~Yang, and H.~Zheng,
\newblock ``The impact of frequency-agility on dynamic spectrum sharing,''
\newblock in {\em IEEE DySPAN}, 2010.

\bibitem{zhang08}
J.~Zhang, D.~Zhang, and M.~Chiang,
\newblock ``The impact of stochastic noisy feedback on distributed network
  utility maximization,''
\newblock {\em IEEE Transactions on Information Theory}, , no. 2, pp. 645--665,
  2008.

\bibitem{basar99}
T.~Basar and G.~Olsder,
\newblock {\em Dynamic Noncooperative Game Theory},
\newblock SIAM, 1999.

\end{thebibliography}

\end{document}